\documentclass[11pt]{article}
\usepackage[T1]{fontenc}
\usepackage{mathtools}
\usepackage{amssymb}
\usepackage{slashed}
\usepackage{fullpage}
\usepackage{accents}
\usepackage[bottom]{footmisc}
\usepackage{mathrsfs}
\usepackage[colorlinks=true,linkcolor=blue,citecolor=magenta,linktocpage=true]{hyperref}
\usepackage[numbers,sort,compress]{natbib}
\usepackage{tocloft}
\usepackage{titlesec}
\usepackage{tikz}
\newcommand{\fL}{\mathfrak{L}}

\def\be#1\ee{\begin{align}#1\end{align}}
\def\bsub#1\esub{\begin{subequations}#1\end{subequations}}

\def\q{\qquad}
\def\f{\frac}
\def\eps{\varepsilon}

\def\lb{\big\lbrace}
\def\rb{\big\rbrace}

\def\oeq{\stackrel{\circ}{=}}

\def\Ju{\tilde{J}}
\def\Jd{\underaccent{\tilde}{J}}
\def\t#1{\tilde{#1}}

\def\de{\mathrm{d}}

\def\A{\mathcal{A}}
\def\B{\mathcal{B}}

\def\I{\mathcal{I}}

\def\L{\mathcal{L}}

\def\O{\mathcal{O}}

\def\Q{\mathcal{Q}}

\numberwithin{equation}{section}

\def\red#1{\textcolor{red}{#1}}

\def\orange#1{\textcolor{orange}{#1}}
\begin{document}
\title{\Large{\textbf{\sffamily Field-dependent diffeomorphisms and\\ the transformation of 
surface charges between gauges}}}
\author{\sffamily Luca Ciambelli${}^1\footnote{\href{mailto:ciambelli.luca@gmail.com}{ciambelli.luca@gmail.com}}$~ \& Marc Geiller${}^2$\footnote{\href{mailto:marc.geiller@ens-lyon.fr}{marc.geiller@ens-lyon.fr}}}
\date{}
\date{\small{\textit{
${}^1$Perimeter Institute for Theoretical Physics,\\
31 Caroline Street North, Waterloo, Ontario, Canada N2L 2Y5\\
${}^2$ENS de Lyon, CNRS, LPENSL, UMR 5672, 69342 Lyon cedex 07, France~\\}}}

\maketitle

\begin{abstract}
When studying gauge theories in the presence of boundaries, local symmetry transformations are typically classified as gauge or physical depending on whether the associated charges vanish or not. Here, we propose that physical charges should further be refined into ``dynamical'' or ``kinematical'' depending on whether they are associated with flux-balance laws or not. To support this proposal, we analyze (A)dS$_3$ gravity with boundary Weyl rescalings and compare the solution spaces in Bondi--Sachs and Fefferman--Graham coordinates. Our results show that the Weyl charge vanishes in the Bondi--Sachs gauge but not in the Fefferman--Graham gauge. Conversely, the charges arising from the metric Chern--Simons Lagrangian behave in the opposite way. This indicates that the gauge-dependent Weyl charge differs fundamentally from charges like mass and angular momentum. This interpretation is reinforced by two key observations: the Weyl conformal factor does not satisfy any flux-balance law, and the associated charge arises from a corner term in the symplectic structure. These properties justify assigning the Weyl charge a kinematical status. These results can also be derived using the field-dependent diffeomorphism that maps between the two gauges. Importantly, this diffeomorphism does not act tensorially on the variational bi-complex due to its field dependency, and is able to “toggle” charges on or off. This provides an example of a large diffeomorphism \textit{between} gauges, as opposed to a residual diffeomorphism \textit{within} a gauge.
\end{abstract}

\thispagestyle{empty}
\newpage
\setcounter{page}{1}
\hfill
\tableofcontents
\bigskip
\hrule

\section{Introduction and motivations}

When studying gauge field theories in the presence of boundaries (be they finite or asymptotic), Noether's theorems imply that local symmetry transformations, or gauge transformations, can give rise to lower-degree conservation laws associated with codimension-two surface\footnote{In three-dimensional spacetimes these ``surface'' charges reduce to one-dimensional ``circle'' charges.} charges \cite{Regge:1974zd,Wald:1993nt,Iyer:1994ys,Barnich:1995ap,Barnich:2000zw,Barnich:2001jy,Barnich:2003xg}. The existence of these surface charges expresses the crucial fact that not all gauge transformations are equivalent. Rather, they can be classified as either ``proper'' or ``improper'' (to use the Hamiltonian terminology of Regge and Teitelboim \cite{Regge:1974zd}) or equivalently as ``pure gauge'' or ``physical'', depending on whether they give rise to vanishing surface charges or not. Transformations with vanishing surface charges represent true gauge redundancies of a theory, while those with non-trivial charges represent physical transformations which map between inequivalent field configurations. In short, one can think of boundaries as being able to breathe life into gauge transformations and potentially turning bulk gauge redundancies into physical symmetries (for a review, see \cite{Ciambelli:2022vot} and references therein). These physical charges can furthermore be endowed with algebraic structures which play a central role in the description of the classical and quantum dynamics of a theory, in particular through the prism of holography.\footnote{We refer the reader to \cite{Harlow:2018tng} for more details and useful references.}

Specifically, a lot of attention has been devoted to the study of asymptotic symmetries and the associated charges in gravitational systems. Historically this has played a seminal role in the inception of AdS/CFT \cite{Brown:1986nw,Coussaert:1995zp}. More recently, in the context of asymptotically-flat spacetimes, surface charges and their algebraic properties have been at the center of interbreeding developments on flat holography \cite{Barnich:2009se,Barnich:2010eb,Strominger:2014pwa}, infrared properties of gravitons \cite{He:2014laa, Cachazo:2014fwa}, and gravitational wave physics \cite{Strominger:2013jfa,Strominger:2014pwa, Pasterski:2015tva,Compere:2019gft,Mitman:2022kwt}. As reviewed by Strominger \cite{Strominger:2017zoo}, this has revealed in particular that the infinite-dimensional BMS asymptotic symmetry group and its charges, related to mass and angular momentum, are intimately related to the degeneracy properties of the asymptotic radiative vacua and to gravitational wave memory observables. While most of this progress has taken place in four-dimensional asymptotically-flat spacetimes, there are also new encouraging results in dS \cite{Anninos:2010zf,Poole:2018koa,PremaBalakrishnan:2019jvz,Compere:2020lrt,Compere:2019bua,Poole:2021avh,Compere:2023ktn,Compere:2024ekl}, and other theories such as QED \cite{He:2014cra, Kapec:2017tkm}, QCD \cite{Pate:2019lpp}, and three-dimensional gravity \cite{Compere:2013bya,Troessaert:2013fma,Perez:2016vqo,Grumiller:2016pqb,Grumiller:2017sjh,Campoleoni:2018ltl,Ojeda:2019xih,Grumiller:2019fmp,Adami:2020ugu,Ciambelli:2020ftk,Ciambelli:2020eba,Alessio:2020ioh,Fiorucci:2020xto,Ruzziconi:2020wrb,Geiller:2021vpg,Adami:2022ktn,Campoleoni:2022wmf,Adami:2023fbm,McNees:2023tus,Ciambelli:2023ott,Arenas-Henriquez:2024ypo} have been extensively used as testbeds and arenas for case studies.

There are two particular and related directions along which recent developments in the study of gravitational charges have unfolded. The first one is the investigation of the various solution spaces and charges which can be defined when progressively relaxing boundary conditions. In four-dimensional spacetimes, this has motivated in particular the study of the extended and generalized BMS groups \cite{Barnich:2010eb,Barnich:2011mi,Flanagan:2015pxa,Compere:2018ylh,Campiglia:2020qvc}, of the BMS--Weyl group \cite{Freidel:2021fxf}, of the so-called partial Bondi gauge \cite{Geiller:2022vto,Geiller:2024amx,McNees:2024iyu}, and of a recent solution space with twist \cite{Mao:2024jpt}. Such extensions of the asymptotic structure were also investigated thoroughly in three-dimensional gravity \cite{Compere:2013bya,Troessaert:2013fma,Perez:2016vqo,Grumiller:2016pqb,Grumiller:2017sjh,Campoleoni:2018ltl,Ojeda:2019xih,Grumiller:2019fmp,Adami:2020ugu,Ciambelli:2020ftk,Ciambelli:2020eba,Alessio:2020ioh,Fiorucci:2020xto,Ruzziconi:2020wrb,Geiller:2021vpg,Adami:2022ktn,Campoleoni:2022wmf,Adami:2023fbm,McNees:2023tus,Ciambelli:2023ott,Arenas-Henriquez:2024ypo}. The second development is the study of the tools which enable to actually identify charges and fluxes, and to discuss properties such as the integrability of the charges and their algebra. This has led to technical developments pertaining to covariant phase space methods, including the analysis of corner ambiguities \cite{Compere:2008us,Geiller:2017xad,DePaoli:2018erh,Oliveri:2019gvm,Freidel:2020xyx,Freidel:2020svx}, of slicing ambiguities \cite{Ruzziconi:2020wrb,Adami:2021nnf}, of the Wald--Zoupas prescription \cite{Wald:1999wa,Chandrasekaran:2021vyu,Grant:2021sxk,Odak:2022ndm,Odak:2023pga,Rignon-Bret:2024gcx}, the introduction of new charge brackets \cite{Barnich:2010eb,Freidel:2021cbc,Freidel:2021dxw, Wieland:2021eth}, and the use of so-called embedding techniques \cite{Ciambelli:2021nmv,Ciambelli:2022cfr,Ciambelli:2021vnn}.

In spite of these impressive technical developments, some properties of surface charges and of the covariant phase space seem (to the best of our knowledge) to have gone mostly unnoticed. Indeed, it appears that relaxing boundary conditions can lead to charges with peculiar properties such as a dependency on the coordinate gauge. Such charges are vanishing in a given gauge, while they can be non-vanishing and therefore physical in another one. Relatedly, it turns out that the diffeomorphism which maps between two given gauges is generically field-dependent, and as such has a non-tensorial action on variational forms. This peculiar transformation behavior arises from the necessary variations of the field-dependent change of coordinates. Meticulously addressing these aspects can elucidate the physical significance and gauge covariance of asymptotic charges. This is particularly pertinent given the recent surge in the discovery of new asymptotic symmetries and charges in the literature.

The aim of the present work is to illustrate these features in what seems to be the simplest example, namely (A)dS$_3$ gravity with boundary conditions allowing for Weyl conformal rescalings of the boundary metric. Weyl charges in three-dimensional gravity have already been studied extensively \cite{Imbimbo:1999bj,Bautier:1999ic,Troessaert:2013fma,Apolo:2014tua,Alessio:2020ioh,Ciambelli:2023ott,Arenas-Henriquez:2024ypo}. These studies were performed in Fefferman--Graham (FG) gauge,\footnote{And the so-called the Weyl--Fefferman--Graham gauge, introduced in \cite{Ciambelli:2019lap} and further studied in \cite{Ciambelli:2023ott,Arenas-Henriquez:2024ypo}.} with the exception of \cite{Ciambelli:2020ftk,Ciambelli:2020eba,Geiller:2021vpg} dealing with the Bondi--Sachs (BS) gauge as well. Comparing in parallel the fate of the Weyl transformations and the associated charges in FG and BS gauge reveals the surprising fact at the origin of the present work, namely that the Weyl charges are vanishing in BS gauge while they are non-vanishing in FG gauge. According to the standard terminology, these charges would therefore be pure gauge in BS gauge but physical in FG gauge. Instead, we propose a finer notion of physical charges, which we call ``kinematical'' or ``dynamical''. The (A)dS$_3$ Weyl charges are examples of kinematical charges in the sense that they can be turned off by changing the coordinate gauge. In addition, there are other salient features which single out the kinematical Weyl charges, namely the fact that they satisfy no flux-balance law, and that they can be seen as arising from a corner contribution to the symplectic structure. As an extra curious result, we compute the charges associated with both the Einstein--Hilbert (EH) and the metric Chern--Simons (CS) Lagrangians \eqref{EH and CS Lagrangians}, to discover that the situation in the metric CS theory is opposite to that in EH gravity: the Weyl charges in the metric CS theory are vanishing in FG gauge but non-vanishing in BS gauge. This illustrates how two formulations of the same theory\footnote{Strictly speaking, the metric CS theory allows for a larger solution space than EH gravity, because its equations of motion are the vanishing of the Cotton tensor instead of the Einstein tensor. Therefore, while solutions of EH gravity are solutions of the metric CS theory, the converse is not true. For our purposes however, since we consider a solution space (the on-shell (A)dS$_3$ metrics with Weyl freedom) derived from EH gravity, it is automatically a solution space for the metric CS theory. We refer the reader to \cite{Adami:2021sko} and references therein.} can have different (kinematical, in this case) charges \cite{DePaoli:2018erh,Oliveri:2019gvm,Freidel:2020xyx,Freidel:2020svx}. Based on these observations, one can conclude that the extra charges which have been identified in the partial Bondi gauge \cite{Geiller:2022vto,Geiller:2024amx} and in various relaxations of the three-dimensional boundary conditions \cite{Troessaert:2013fma,Grumiller:2016pqb,Ruzziconi:2020wrb,Adami:2020ugu,Alessio:2020ioh,Geiller:2021vpg,Adami:2022ktn,Ciambelli:2023ott,Arenas-Henriquez:2024ypo} provide further examples of kinematical charges.

Aside from these results on the properties of asymptotic charges, we touch upon an important technical subtlety of the covariant phase space formalism \cite{GawEdzki1972,Kijowski1973,Kijowski1976,Crnkovic1986, Crnkovic1988} and the variational bi-complex \cite{Vinogradov1977,Vinogradov1978,Vinogradov1984,Vinogradov1984a,Tulczyjew1980,MR1188434,Anderson:1996sc,Barnich:2001jy}, which is that forms on field-space do not transform as tensors under the diffeomorphisms which map between gauges. The reason for this is that such diffeomorphisms are typically field-dependent, and that this field-dependency percolates into the change of coordinates. This means that when expressing the old coordinates in terms of the new ones, the field-dependency implies that $\delta\tilde{x}(x)\neq0$ and $[\tilde{\partial}_\mu,\delta]\neq0$. Just like the ordinary bracket of vector fields must be adjusted as in \eqref{modified bracket} in order to take into account the field-dependency typically present in asymptotic Killing vectors \cite{Barnich:2010eb}, the transformation of the asymptotic Killing vectors between the gauges must be corrected by a variational term as well \cite{Compere:2016hzt}. The resulting formula is
\be
\xi^\mu=\Jd^\mu_\alpha\tilde{\xi}^\alpha+\Jd^\mu_\alpha\fL_\xi\tilde{x}^\alpha,
\ee
where $\xi$ and $\tilde{\xi}$ are the vector fields in the coordinate gauges $x$ and $\tilde{x}$, $\Jd$ is the Jacobian, and $\fL_\xi$ is the field-space Lie derivative which inputs the transformation laws of the fields entering the expression $\tilde{x}(x)$. Using similar formulas, one can derive the transformation behavior of the local pre-symplectic potential. We find that it is given by
\be\label{transformation theta intro}
\Theta^\mu=|\Ju|\Jd^\mu_\alpha\tilde{\Theta}^\alpha-\Jd^\mu_\alpha\A^\alpha,
\ee
where the correction term $\A$ given by \eqref{A anomaly} depends solely on the variations of the coordinates $\delta\tilde{x}(x)$. Interestingly, the example of (A)dS$_3$ with Weyl freedom is simple enough to allow for these structures to be computed and verified explicitly. In particular, \eqref{transformation theta intro} contains two sources of complications. The first one is the extra variational term $\A$, which has the complicated expression \eqref{A anomaly}. The second one is the fact that $\tilde{\Theta}$ on the right-hand side must be computed while taking into account the fact that $\delta\tilde{x}(x)\neq0$.

As one of the results, we study in detail the field-dependent diffeomorphism between the BS and FG gauges \cite{Poole:2018koa,Compere:2019bua,Ciambelli:2020ftk}. This enables us to confirm what has been observed at the level of the charges, namely that for EH gravity the Weyl charge vanishes in BS gauge while it is non-zero in FG gauge. We also show that the FG gauge with non-vanishing Weyl factor in the metric can be reached from the BS gauge with vanishing Weyl factor. This allows us to conclude that it is really the field-dependency of the diffeomorphism between the BS and FG gauges which turns on the Weyl charge: in this sense, this diffeomorphism is large. These computations raise important questions concerning the very foundations of the covariant phase space formalism and of the variational bi-complex. Indeed, non-tensorial transformations laws such as \eqref{transformation theta intro} can also be derived for objects such as the Iyer--Wald \cite{Iyer:1994ys} or the Barnich--Brandt \cite{Barnich:2001jy} two-forms used to compute the charges. The question is then that of the proper mathematical formalism to take into account the fact that field-space forms transform as connections rather than tensors.

Below is a table summarizing the results of this work and presenting  the outline and content of the various section. We begin in section \ref{sec:BS} with a study of the solution space and the asymptotic symmetries and charges in Bondi--Sachs (BS) gauge. This gauge is more commonly used to study asymptotically-flat spacetimes and the three-dimensional BMS$_3$ group \cite{Ashtekar:1996cd,Barnich:2006av}, but it can also handle a non-vanishing cosmological constant \cite{Poole:2018koa,Compere:2020lrt,Compere:2019bua,Geiller:2021vpg,Geiller:2022vto}. We present the solution space allowing for Weyl rescaling in AdS$_3$, and then study the residual diffeomorphisms. These contain a three-dimensional freedom of performing the double Witt and Weyl transformations. We then compute the asymptotic charges to discover, as announced, that the Weyl charge vanishes in EH gravity but not in the metric CS theory. In section \ref{sec:FG} we then mimic exactly the structure of section \ref{sec:BS}, but working this time in Fefferman--Graham (FG) gauge. In particular, it is possible to find field redefinitions which put the BS and FG results in very similar form. This allows us to compare the two gauges. In particular, the crucial difference is that in FG gauge the Weyl charge is non-vanishing for EH gravity but vanishing for the metric CS theory. We interpret this different behavior of the Weyl charge in BS and FG gauge as a characteristic of a kinematical charge, as opposed to mass and angular momentum which cannot be removed by a change of gauge. We also show in BS and FG gauge that the Weyl charge arises from a corner term in the symplectic structure. Section \ref{sec:diffeo} is devoted to the study of the diffeomorphism between the BS and FG gauges. We explain how this diffeomorphism is constructed perturbatively by mapping the metric, and then use it to map the asymptotic Killing vectors as well as the pre-symplectic potentials. In the case of the Killing vectors, one can clearly see that it is the field-dependency of the diffeomorphism which enables us to transport the Weyl rescaling symmetry from the BS to the FG gauge. Neglecting this field-dependency, one would wrongly go from a BS gauge with Weyl symmetry to an FG gauge without Weyl symmetry. Finally, we give our conclusions and perspectives for future work in section \ref{sec:conclusion}.

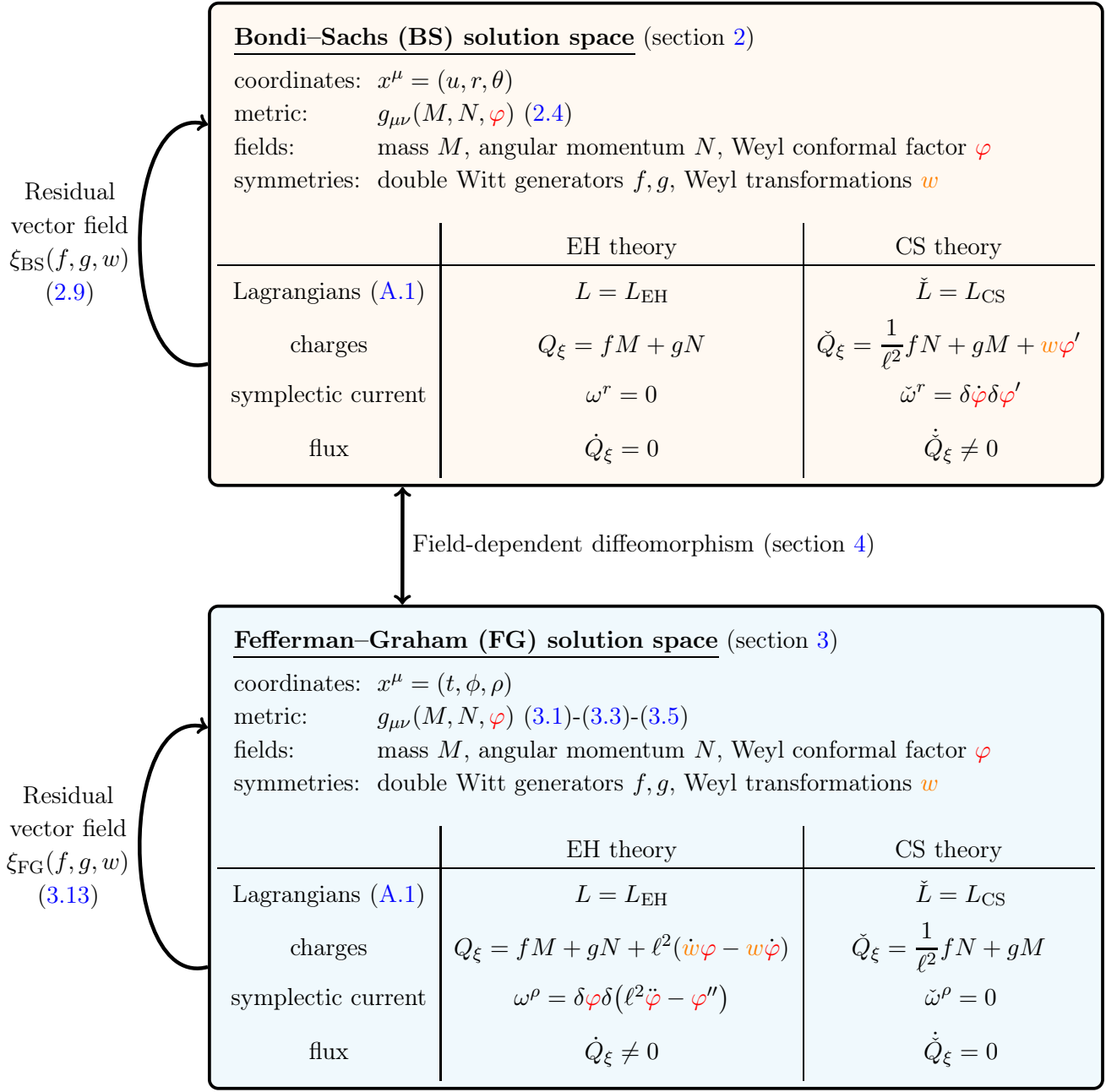
\begin{center}
\hspace{-1cm}
\begin{tikzpicture}
\node (A) [draw, fill=orange!5, line width=0.5mm, rectangle, rounded corners, minimum height=4cm, minimum width=4cm, align=left] at (0,0) {$\displaystyle\vphantom{\int}$$\phantom{\int}$\textbf{\underline{Bondi--Sachs (BS) solution space}} (section \ref{sec:BS})
\\$\phantom{\int}$coordinates:\phantom{symmetries:metric:fields:}\hspace{-4cm} $x^\mu=(u,r,\theta)$
\\$\phantom{\int}$metric:\phantom{symmetries:coordinates:fields:}\hspace{-4cm} $g_{\mu\nu}(M,N,\red{\varphi})$ \eqref{BS-Weyl line element}
\\$\phantom{\int}$fields:\phantom{symmetries:coordinates:metric:}\hspace{-4cm} mass $M$, angular momentum $N$, Weyl conformal factor $\red{\varphi}$
\\$\phantom{\int}$symmetries:\phantom{coordinates:metric:fields:}\hspace{-4cm} double Witt generators $f,g$, Weyl transformations $\orange{w}$
\\~\\
$$\begin{tabular}{c|c|c} 
& \vspace{-0.4cm}\phantom{$Q_\xi=fM+gN+\ell^2(\dot{w}\red{\varphi}-w\dot{\red{\varphi}})$} &\\
& EH theory & CS theory\\\hline
Lagrangians \eqref{EH and CS Lagrangians} & $L=L_\text{EH}$ & $\displaystyle\phantom{\int}\check{L}=L_\text{CS}$\\ 
charges & $Q_\xi=fM+gN$ & $\displaystyle\check{Q}_\xi=\f{1}{\ell^2}fN+gM+\orange{w}\red{\varphi}'$\\
symplectic current & $\omega^r=0$ & $\displaystyle\phantom{\int}\check{\omega}^r=\delta\dot{\red{\varphi}}\delta\red{\varphi}'$\\
flux & $\dot{Q}_\xi=0$ & $\displaystyle\phantom{\int}\dot{\check{Q}}_\xi\neq0$\\
\end{tabular}$$};
\node (B) [draw, fill=cyan!5, line width=0.5mm, rectangle, rounded corners, minimum height=4cm, minimum width=4cm, align=left] at (0,-9.5) {$\displaystyle\vphantom{\int}$$\phantom{\int}$\textbf{\underline{Fefferman--Graham (FG) solution space}} (section \ref{sec:FG})
\\$\phantom{\int}$coordinates:\phantom{symmetries:metric:fields:}\hspace{-4cm} $x^\mu=(t,\phi,\rho)$
\\$\phantom{\int}$metric:\phantom{symmetries:coordinates:fields:}\hspace{-4cm} $g_{\mu\nu}(M,N,\red{\varphi})$ \eqref{general FG metric}-\eqref{FG gamma}-\eqref{conformal gauge}
\\$\phantom{\int}$fields:\phantom{symmetries:coordinates:metric:}\hspace{-4cm} mass $M$, angular momentum $N$, Weyl conformal factor $\red{\varphi}$
\\$\phantom{\int}$symmetries:\phantom{coordinates:metric:fields:}\hspace{-4cm} double Witt generators $f,g$, Weyl transformations $\orange{w}$
\\~\\
$$\begin{tabular}{c|c|c} 
& & \vspace{-0.4cm}\phantom{$\,\check{Q}_\xi=\f{1}{\ell^2}fN+gM+w\red{\varphi}'$}\\
& EH theory & CS theory\\\hline
Lagrangians \eqref{EH and CS Lagrangians} & $L=L_\text{EH}$ & $\displaystyle\phantom{\int}\check{L}=L_\text{CS}$\\ 
charges & $Q_\xi=fM+gN+\ell^2(\dot{\orange{w}}\red{\varphi}-\orange{w}\dot{\red{\varphi}})$ & $\displaystyle\check{Q}_\xi=\f{1}{\ell^2}fN+gM$\\
symplectic current & $\omega^\rho=\delta\red{\varphi}\delta\big(\ell^2\ddot{\red{\varphi}}-\red{\varphi}''\big)$ & $\displaystyle\phantom{\int}\check{\omega}^\rho=0$\\
flux & $\dot{Q}_\xi\neq0$ & $\displaystyle\phantom{\int}\dot{\check{Q}}_\xi=0$\\
\end{tabular}$$};
\draw[<->, line width=0.6mm] ([xshift=8em]A.south west) -- ([xshift=8em]B.north west) node[midway, right, align=center] {Field-dependent diffeomorphism (section \ref{sec:diffeo})};
\path (A) edge [->, line width=0.6mm, loop left,looseness=1] node[align=center] {Residual\\ vector field\\$\xi_\text{BS}(f,g,w)$\\ \eqref{BS vector field}} (A);
\path (B) edge [->, line width=0.6mm, loop left,looseness=1] node[align=center] {Residual\\ vector field\\$\xi_\text{FG}(f,g,w)$\\ \eqref{FG AKV gauge conditions}} (B);
\end{tikzpicture}
\end{center}

Our notations and conventions are as follows. We denote the spacetime Lie derivative by \hbox{$\L_\xi=i_\xi\de+\de i_\xi$} and the field-space Lie derivative by $\fL_\xi=I_\xi\delta+\delta I_\xi$. We denote by $\epsilon$ the Levi--Civita symbol with $\epsilon^{ur\theta}=+1=\epsilon^{t\phi\rho}=-\epsilon^{t\rho\phi}$, and by $\eps$ the Levi--Civita tensor. Anti-symmetrization is $[\mu\nu]=\f{1}{2}(\mu\nu-\nu\mu)$. We use the notation $\oeq$ to denote equalities which hold up to total derivative terms $(\dots)'$ in the angular variable $\theta$ (in BS gauge) or $\phi$ (in FG gauge). This allows us to work with densities without having to integrate them over the celestial circle, while still permitting integration by parts and to freely discard boundary terms.

\section{Bondi--Sachs gauge}
\label{sec:BS}

The example which illustrates the present work arises from studying the fate of the Weyl charges in (A)dS$_3$ gravity using two different gauge fixings for the metric: the Bondi--Sachs (BS) gauge and the Fefferman--Graham (FG) gauge. In both cases, we  gather the various ingredients of the calculation and proceed in a similar way. We start by recalling the gauge fixing conditions and the properties of the on-shell metrics in the presence of a Weyl (or conformal) factor. We then study the asymptotic Killing vectors generating the residual diffeomorphisms. Finally, we compute the codimension-two charges and their algebra in the case of the Einstein--Hilbert (EH) and the metric Chern--Simons (CS) Lagrangians. We begin by performing this study in BS gauge.

\subsection{Gauge fixing and solution space}

Let us consider coordinates $x^\mu=(u,r,\theta)$ and the line element
\be\label{BS line element}
\de s^2=\left(2M-\f{r^2}{\ell^2}\right)\de u^2-2\de u\,\de r+2N\de u\,\de\theta+r^2\de\theta^2,
\ee
which satisfies the three BS gauge conditions
\be\label{BS gauge conditions}
g_{\theta\theta}=r^2,
\q
g_{rr}=0, \q g_{r\theta}=0.
\ee
In these coordinates, the conformal boundary $\B$ is located at $r=+\infty$. The functions $M$ and $N$ depend only on $(u,\theta)$, and will be referred to as the mass and angular momentum aspects. In order for the above metric to solve the (A)dS Einstein equations\footnote{Since we work with $\Lambda=-1/\ell^2$, all the equations are written here for AdS, but the analytic continuation to dS is always possible and does not alter any of the results. An advantage of the BS gauge is that it allows us to also take the flat limit $\ell\to\infty$ at any step. This is however not possible in FG gauge, as can be seen on the line element \eqref{general FG metric}.} $G_{\mu\nu}=\ell^{-2}g_{\mu\nu}$, these two functions must satisfy the evolution equations
\be\label{BS evolution of M and N}
\dot{M}=\f{1}{\ell^2}N',
\q\q
\dot{N}=M',
\ee
where a dot denotes $\partial_u$ and a prime denotes $\partial_\theta$. These flux-balance laws arise respectively from the $uu$ and $u\theta$ Einstein equations. Importantly, this solution of the Einstein field equations is also a solution of the equations of motion arising from the metric CS Lagrangian $\check{L}$ in \eqref{EH and CS Lagrangians}, which amount to the vanishing of the Cotton tensor. This is what will enable us to also consider the charges arising from the BS and FG solution spaces in this metric CS theory.

Let us now consider a finite diffeomorphism which redefines the radial coordinate as $r\mapsto e^{\varphi(u,\theta)}r$, while manifestly preserving the range of all the initial coordinates. This leads to the new line element
\be\label{BS-Weyl line element}
\de s^2=\left(2M-2r\dot{\varphi}e^\varphi-e^{2\varphi}\f{r^2}{\ell^2}\right)\de u^2-2e^\varphi\de u\,\de r+2(N-r\varphi'e^\varphi)\,\de u\,\de\theta+e^{2\varphi}r^2\de\theta^2,
\ee
which is once again a solution of the Einstein equations provided the evolution equations \eqref{BS evolution of M and N} are imposed. This is guaranteed by the fact that we have obtained \eqref{BS-Weyl line element} by performing a diffeomorphism of the solution \eqref{BS line element}, but it can also be verified independently. Importantly, one should note that the new field $\varphi(u,\theta)$ entering the metric \eqref{BS-Weyl line element} has an arbitrary time dependency which is not fixed by any of the field equations. This is in stark contrast with the mass and angular momentum, whose time evolution is constrained by \eqref{BS evolution of M and N}. As customary, we refer to these constrained time evolutions as a \textit{flux-balance laws}. Therefore, the field $\varphi$ does not satisfy any flux-balance law. This already hints at the fact that the fields $\varphi$ and $(M,N)$ labelling the solutions have a different status.

There are of course known situations in which fields with unspecified time dependency enter the metric. This is for example the case of the shear in four-dimensional asymptotically-flat spacetimes, which carries the two degrees of freedom of gravitational radiation. Its time dependency is unspecified by the Einstein equations because its $u$ profile along future null infinity $\I^+$ is a physical information about the radiation of the system, which must be provided in order to reconstruct the solution. In the present situation however, $\varphi$ cannot have this interpretation because the theory is topological and therefore does not admit propagating degrees of freedom. Moreover, at the difference with the shear, here $\varphi$ comes with its own associated (and arbitrarily time-dependent) symmetry parameter, as we will see below.

For the rest of this section, we will study the solution space described by the metric \eqref{BS-Weyl line element} and the three fields $(M,N,\varphi)$. Strictly speaking, this metric is not written in the BS gauge \eqref{BS gauge conditions} anymore (because of the diffeomorphism which we have performed), but satisfies instead the differential determinant condition
\be\label{relaxed determinant}
\partial_r\left(\f{g_{\theta\theta}}{r^2}\right)=0
\ee
initially introduced by Barnich and Troessaert in its four dimensional version \cite{Barnich:2010eb}. In a slight abuse of language, we still refer to \eqref{BS-Weyl line element} as a line element in BS gauge. The main point is that it features, in addition to the mass and angular momentum, the Weyl factor $\varphi$. Indeed, one can compute the induced boundary metric on $\B$ to find
\be
\lim_{r\to\infty}\f{\de s^2}{r^2}=e^{2\varphi}\left(-\f{\de u^2}{\ell^2}+\de\theta^2\right),
\ee
which confirms that $\varphi$ plays the role of the boundary conformal factor. In other words, the radial rescaling through which we have introduced $\varphi$ has induced a Weyl transformation of the boundary metric. 

Using the on-shell line element \eqref{BS-Weyl line element}, we can now compute the radial components of the symplectic potentials \eqref{potentials}, obtained respectively from the variations of the EH and the metric CS Lagrangians. Using the notation $\oeq$ to denote equalities which hold up to total angular derivatives $(\dots)'$ in the angular variable $\theta$, we find
\be\label{theta BS}
\Theta^r\oeq\delta\left(-\f{r^2}{\ell^2}e^{2\varphi}-\f{r}{2}e^\varphi\dot{\varphi}+M\right)+\O(r^{-1}),
\q\q
\check{\Theta}^r\oeq\f{1}{2}\partial_u\big(\delta\varphi'\varphi\big)+\O(r^{-1}).
\ee
The corresponding symplectic current densities are therefore\footnote{Our convention is that field-space variations are anti-commuting.}
\be\label{BS symplectic currents}
\omega^r=\lim_{r\to\infty}\delta\Theta^r=0,
\q\q
\check{\omega}^r=\lim_{r\to\infty}\delta\check{\Theta}^r=-\f{1}{2}\partial_u\big(\delta\varphi'\delta\varphi\big)\oeq\delta\dot{\varphi}\delta\varphi'.
\ee
We will use this result below to compute the symplectic flux and show that it corresponds as expected to the time evolution of the charges. At the difference with the four-dimensional BMS--Weyl framework \cite{Freidel:2021fxf}, in three dimensions the symplectic structure is finite. The charges will therefore be finite as well, and no renormalization is required.

It should be noted that $\varphi$ enters $\check{\Theta}^r$ and $\check{\omega}^r$ as a simple corner contribution $\partial_u(\dots)$, and we will show below that the Weyl contribution to the CS charge can be removed with a corner term. Such a relationship between corner contributions to the symplectic structure and kinematical charges was already alluded to in \cite{McNees:2024iyu}. Similarly, we will see that in FG gauge the Weyl factor enters the symplectic structure of the EH Lagrangian via a corner term.

\subsection{Residual diffeomorphisms}
\label{residual BS}

We now look for the asymptotic Killing vectors $\xi^\mu=(\xi^u,\xi^r,\xi^\theta)$ which preserve the two BS gauge conditions $g_{rr}=0=g_{r\theta}$ as well as the differential determinant condition \eqref{relaxed determinant}. Denoting $\L_\xi$ the spacetime Lie derivative, we find
\bsub\label{BS vector field}
\be
\L_\xi g_{rr}=0\q&\Rightarrow\q\xi^u=f,\\
\L_\xi g_{r\theta}=0\q&\Rightarrow\q\xi^\theta=g-\f{1}{r}e^{-\varphi}f',\\
\partial_r\left(\f{\L_\xi g_{\theta\theta}}{r^2}\right)=0\q&\Rightarrow\q\xi^r=hr+e^{-\varphi}(f''+f'\varphi')-\f{1}{r}e^{-2\varphi}Nf',
\ee
\esub
where $f(u,\theta)$, $g(u,\theta)$, and $h(u,\theta)$ are the three symmetry generators. The last condition is the preservation of the differential determinant gauge condition \eqref{relaxed determinant}, which enables to have $\varphi\neq0$ and the associated symmetry parameter $h$. In addition to preserving the gauge conditions, we also need to preserve the fall-offs by requiring
\bsub\label{BS conditions on f and g}
\be
\L_\xi g_{u\theta}=\O(r)\q&\Rightarrow\q\dot{g}=\f{1}{\ell^2}f',\\
\L_\xi g_{ur}+\f{e^{-\varphi}}{2r^2}\L_\xi g_{\theta\theta}=0\q&\Rightarrow\q\dot{f}=g',
\ee
\esub
which manifestly fixes the time-dependency of the symmetry generators $f$ and $g$. At this stage, one should recall that symmetry generators are in correspondence with fields in the solution space\footnote{More precisely, the asymptotic charges can be viewed as pairings between symmetry generators and fields of the solution space. There exist however cases where there are more fields satisfying flux-balance laws than parameters in the residual diffeomorphism. This is the case in four-dimensional asymptotically-flat spacetimes, where the fields $E^n_{ab}$ appearing at order $r^{-n}$ in the transverse metric have no corresponding symmetry generators \cite{Geiller:2024bgf}. However, these fields do not appear in the asymptotic charges.}. Here, the pairing is between $(f,M)$, $(g,N)$, and $(h,\varphi)$. Since the fields $M$ and $N$ satisfy evolution equations, the parameters do so as well. Similarly, since the Weyl factor $\varphi$ has an arbitrary time dependency (i.e., no flux-balance law), its associated symmetry parameter has an unconstrained time dependency as well. In fact \cite{Alessio:2020ioh}, it will prove useful to consider a field-dependent redefinition of this parameter (sometimes called ``change of slicing'' \cite{Ruzziconi:2020wrb,Geiller:2021vpg,Adami:2021nnf}), obtained by setting
\be\label{BS Weyl slicing}
h=w-f\dot{\varphi}-g\varphi'-g',
\ee
where the new free function is now $w(u,\theta)$. This will guarantee later on that the charges coming from the metric CS term are integrable, and also lead to a very simple action \eqref{BS transformation law of Weyl} of the Weyl transformations on the field $\varphi$.

With the residual diffeomorphism at our disposal, we can now compute its action on the solution space to find that the various fields transform as
\bsub\label{BS transformation laws}
\be
\fL_\xi M&=f\dot{M}+gM'+2g'M+\f{2}{\ell^2}f'N-g''',\\
\fL_\xi N&=f\dot{N}+gN'+2g'N'+2f'M-f''',\\
\fL_\xi\varphi&=w,\label{BS transformation law of Weyl}
\ee
\esub
where $\fL_\xi=I_\xi\delta+\delta I_\xi$ is the field-space Lie derivative and $I_\xi$ the field-space interior product.\footnote{We are adapting the conventions of \cite{Ciambelli:2022vot}.} One can note that setting $\varphi=0$ constrains $w$ to be vanishing identically without leaving the possibility of zero modes, while on the other hand setting $M$ or $N$ to zero leaves a three-dimensional freedom in $g$ and $f$ respectively, which corresponds to the global isometries. In this sense, the Weyl symmetry is directly tied to the presence of the field $\varphi$ in the solution space.

In order to compute the algebra of the vector fields, we use the modified bracket \cite{Barnich:2010eb} which takes into account the field-dependency of the residual Killing vectors,
\be\label{modified bracket}
\big[\xi_1,\xi_2\big]_*=\big[\xi_1,\xi_2\big]-\fL_{\xi_1}\xi_2+\fL_{\xi_2}\xi_1=\xi(f_{12},g_{12},w_{12}).
\ee
In the case $\fL_{\xi}f=0$, $\fL_{\xi}g=0$, and $\fL_{\xi}w=0$, we obtain
\bsub\label{BS algebra of vectors}
\be
f_{12}&=f_1g'_2+g_1f'_2
-(1\leftrightarrow2),\\
g_{12}&=\f{1}{\ell^2}f_1f'_2+g_1g'_2
-(1\leftrightarrow2),\\
w_{12}&=0.
\ee
\esub
This is the double Witt algebra\footnote{More precisely, the two commuting copies of Witt can be obtained by considering chiral combinations of the parameters $f$ and $g$ \cite{Barnich:2012aw}.} with an extra abelian $\mathbb{R}$ component arising from the Weyl transformations.

\subsection{Charges}

We can now compute the charges at cuts of the asymptotic boundary at fixed time $u$ and large radius. For this, we use the Iyer--Wald expressions given in \eqref{charge aspects} and derived from the EH and metric CS Lagrangians.\footnote{We checked that these charges coincide with the Barnich--Brandt charges obtained from cohomological methods \cite{Barnich:2001jy}, since the extra term relating them vanishes in our setup.} Computing the component $\mu,\nu=u,r$, we find that the charges in both cases are integrable and given by
\be\label{BS charges}
Q_\xi=\lim_{r\to\infty}\Q_\xi^{ur}\oeq fM+gN,
\q\q
\check{Q}_\xi=\lim_{r\to\infty}\check{\Q}_\xi^{ur}\oeq\f{1}{\ell^2}fN+gM+w\varphi'.
\ee
In this calculation, one should note that all the terms in the Iyer--Wald formulas \eqref{charge aspects} contribute non-trivially. One can also notice that
\be\label{BS duality}
Q_\xi=\check{Q}_\xi\left(f\to g\,,\,g\to\f{f}{\ell^2}\,,\,w\to0\right),
\ee
which reveals (outside of the Weyl sector) a duality between the EH and metric CS theories. This duality was also revealed in first order connection-triad variables in \cite{Geiller:2020edh,Geiller:2020okp}. One should note that the integrability of the CS charge is made possible by the change of slicing \eqref{BS Weyl slicing} for the Weyl transformations.

Importantly, it is possible to show that the Weyl contribution to $\check{Q}_\xi$ can be seen as arising from a corner symplectic potential $\vartheta$. Indeed, adding to the symplectic two-form the term $\text{d}\delta\vartheta$ with $\vartheta=-\f{1}{2}\varphi'\delta\varphi$, one finds that its contribution to the charge is
\be
I_\xi(\delta\vartheta)=-\f{1}{2}I_\xi(\delta\varphi'\delta\varphi)=-\f{1}{2}\delta(w'\varphi-w\varphi')\oeq\delta(w\varphi'),
\ee
which is precisely the Weyl term in $\check{Q}_\xi$ above. If one chooses the viewpoint that the corner term is an ambiguity which can be played with at will, a corner term proportional to $\vartheta$ can be added to the symplectic two-forms $\omega$ or $\check{\omega}$, and the Weyl charge can be introduced or removed from $Q_\xi$ or $\check{Q}_\xi$ in an arbitrary way. It is therefore puzzling that, although according to the usual classification the Weyl charge should be considered as physical for the metric CS theory, one can eliminate it by the addition of a corner term. This is obviously not possible for the mass and angular momentum charges, which therefore reinforces the idea that $(M,N)$ and $\varphi$ are indeed on a different footing. We propose to label the charges associated with the fields $(M,N)$ as ``dynamical'' physical charges, while the charge associated with $\varphi$ is a ``kinematical'' physical charge.

We can now compute the time evolution of the charges to obtain the associated fluxes. Using the evolution equations \eqref{BS evolution of M and N} and the conditions \eqref{BS conditions on f and g}, we find that the fluxes are given by
\be
\dot{Q}_\xi\oeq0,
\q\q
\dot{\check{Q}}_\xi\oeq\dot{w}\varphi'-w'\dot{\varphi}.
\ee
As expected, the lack of conservation of the charge $\check{Q}_\xi$ can be traced back to the arbitrary time dependency in the Weyl factor $\varphi$ and in the associated symmetry parameter $w$. Consistently, one can also check that this time evolution is the one derived from the symplectic flux. Indeed, using \eqref{BS symplectic currents} and \eqref{BS transformation laws} we find
\be\label{BS symplectic flux}
\delta\dot{Q}_\xi\oeq I_\xi\omega^r,
\q\q
\delta\dot{\check{Q}}_\xi\oeq I_\xi\check{\omega}^r.
\ee
Finally, we can compute the algebra of the charges obtained in the EH and metric CS theories. These charge algebras are
\bsub\label{BS charge algebra}
\be
\lb Q_{\xi_1},Q_{\xi_2}\rb=\fL_{\xi_2}Q_{\xi_1}&\oeq Q_{[\xi_1,\xi_2]_*}-f_1g_2'''-(1\leftrightarrow2),\\
\lb\check{Q}_{\xi_1},\check{Q}_{\xi_2}\rb=\fL_{\xi_2}\check{Q}_{\xi_1}&\oeq\check{Q}_{[\xi_1,\xi_2]_*}-\f{1}{2}\left(\f{1}{\ell^2}f_1f_2'''+g_1g_2'''-w_1w_2'\right)-(1\leftrightarrow2),
\ee
\esub
and we therefore find two different central extensions for the two charges, with that for $\check{Q}_\xi$ actually involving the Weyl symmetry parameter. Since in \eqref{BS algebra of vectors} we found that the Weyl symmetry parameters were abelian, the presence of a central extension in \eqref{BS charge algebra} indicates that the Weyl charges actually form a Heisenberg algebra. The total charge algebra is therefore double Virasoro for the EH Lagrangian, and double Virasoro in direct product with Heisenberg in the case of the CS Lagrangian. 

\section{Fefferman--Graham gauge}
\label{sec:FG}

We are now ready to repeat the steps of the previous section, but this time using the FG gauge instead of the BS one. This will reveal in particular that in FG gauge the Weyl charge is non-vanishing for the EH Lagrangian but absent in the metric CS theory, which is the opposite of what we have obtained above in BS gauge. For the EH part this construction in FG gauge is similar to that of \cite{Troessaert:2013fma, Alessio:2020ioh} (see as well \cite{Ciambelli:2020ftk, Ciambelli:2020eba}), although we present it here in a more streamlined manner and by emphasizing the similarities and differences with the BS gauge.

\subsection{Gauge fixing and solution space}
\label{sec:FG solution}

Let us consider coordinates $x^\mu=(\rho,x^a)$ with $x^a=(t,\phi)$ and the line element
\be\label{general FG metric}
\de s^2=\f{\ell^2}{\rho^2}\de\rho^2+\gamma_{ab}\,\de x^a\,\de x^b,
\ee
which satisfies the three FG gauge conditions
\be\label{FG gauge conditions}
g_{\rho\rho}=\f{\ell^2}{\rho^2},
\q
g_{\rho t}=0, \q g_{\rho\phi}=0.
\ee
In these coordinates, the conformal boundary $\B$ is located at $\rho=0$. The Einstein field equations $G_{\mu\nu}=\ell^{-2}g_{\mu\nu}$ are solved by the finite radial expansion
\be\label{FG gamma}
\gamma_{ab}(\rho,x^a)=\f{1}{\rho^2}g_{ab}^{(0)}(x^a)+g_{ab}^{(2)}(x^a)+\rho^2g_{ab}^{(4)}(x^a),
\ee
with
\be\label{on-shell FG conditions}
g_{ab}^{(4)}=\f{1}{4}g_{ac}^{(2)}g^{cd}_{(0)}g_{db}^{(2)},
\q\q
g^{ab}_{(0)}g_{ab}^{(2)}=-\f{\ell^2}{2}R^{(0)},
\q\q
D^a_{(0)}g_{ab}^{(2)}=-\f{\ell^2}{2}\partial_bR^{(0)}.
\ee
We now supplement the FG gauge condition by a conformal gauge fixing for the boundary metric, which we write as
\be\label{conformal gauge}
g_{ab}^{(0)}=e^{2\varphi(x^a)}
\begin{pmatrix}
-1/\ell^2&0\\
0&1
\end{pmatrix}.
\ee
With this conformal gauge we have $R^{(0)}=2e^{-2\varphi}(\ell^2\ddot{\varphi}-\varphi'')$, where a dot denotes $\partial_t$ and a prime denotes $\partial_\phi$. The second condition in \eqref{on-shell FG conditions} can be rewritten as
\be
g_{\phi\phi}^{(2)}=\ell^2\left(g_{tt}^{(2)}-\f{1}{2}e^{2\varphi}R^{(0)}\right).
\ee
Then, the two equations in the third condition \eqref{on-shell FG conditions} can be written as constraints on the time evolution of $g_{ab}^{(2)}$. The components $t$ and $\phi$ are solved respectively by imposing
\be
\dot{g}_{tt}^{(2)}=\f{1}{\ell^2}\big(g_{t\phi}^{(2)}\big)'+\dot{\varphi}(\varphi''-\ell^2\ddot{\varphi})+\ell^2\dddot{\varphi}-\dot{\varphi}'',
\q\q
\dot{g}_{t\phi}^{(2)}=\big(g_{tt}^{(2)}\big)'+\varphi'(\varphi''-\ell^2\ddot{\varphi}).
\ee
Since our goal is to draw a parallel between the computation of the charges in BS and FG gauge, it is very convenient to trade the two free functions in $g_{ab}^{(2)}$ for $M$ and $N$ by defining
\be
M\coloneqq g_{tt}^{(2)}+\f{\ell^2}{2}(\dot{\varphi}^2-2\ddot{\varphi})+\f{1}{2}(\varphi')^2,
\q\q
N\coloneqq g_{t\phi}^{(2)}+\ell^2(\dot{\varphi}\varphi'-\dot{\varphi}').
\ee
In terms of this new data, the evolution equations take the same form as in BS gauge, namely
\be\label{FG evolution of M and N}
\dot{M}=\f{1}{\ell^2}N',
\q\q
\dot{N}=M'.
\ee
We now have a complete characterization of the solution space in FG gauge with conformal boundary metric. It has an arbitrary function $\varphi(x^a)$ with an unconstrained time dependency, and two functions $M(x^a)$ and $N(x^a)$ subject to the above temporal evolution equations.

As a side remark, it should be noted that here we have introduced the Weyl factor directly with our choice of conformally-flat boundary metric \eqref{conformal gauge}, whereas in BS gauge we have introduced it by performing the diffeomorphism $r\mapsto e^{\varphi(u,\theta)}r$ on the line element \eqref{BS line element}. One can therefore wonder whether it is possible to obtain the FG line element with $\varphi\neq0$ from a diffeomorphism of the FG line element with $\varphi=0$. This is indeed possible with the perturbative diffeomorphism
\be
t\mapsto t+\sum_{n=2}^\infty\rho^nT_n(x^a),
\q\q
\phi\mapsto\phi+\sum_{n=2}^\infty\rho^nF_n(x^a),
\q\q
\rho\mapsto e^{-\varphi}\rho+\sum_{n=3}^\infty\rho^nR_n(x^a),
\ee
where the functions are determined by the requirement of preserving the gauge conditions \eqref{FG gauge conditions}. In particular, $R_n$ is determined by the preservation of $g_{\rho\rho}=\ell^2/\rho^2$ and the solution at leading order starts with $R_3=e^\varphi\ell^{-4}\big(T_2^2-\ell^2F_2^2\big)$. Preserving $g_{\rho t}=0$ and $g_{\rho\phi}=0$ at leading order then determines respectively $T_2$ and $F_2$, and one can iterate the expansion. This way of introducing the Weyl factor in FG gauge is obviously more cumbersome and less elegant than by simply inputing \eqref{conformal gauge}. Indeed, as noted in \cite{Ciambelli:2019bzz}, to preserve the FG gauge rescalings of the radial coordinate must be accompanied by subleading transformations of the boundary coordinates. Using radial rescalings to induce a different representative in the conformal class of boundary metrics is a well-known procedure in AdS/CFT, and it relates to the holographic computation of the Weyl anomaly \cite{Henningson:1998gx, Mazur:2001aa, Skenderis:2002wp, Ciambelli:2019bzz}.

Going back to the on-shell line element \eqref{general FG metric}, we can now compute the radial component of the symplectic potentials \eqref{potentials}. We find
\be\label{theta FG}
\Theta^\rho\oeq\delta\left(\f{e^{2\varphi}}{\ell^2\rho^2}+\f{\ell^2}{2}\ddot{\varphi}\right)+\delta\varphi\big(\varphi''-\ell^2\ddot{\varphi}\big)+\O(\rho),
\q\q
\check{\Theta}^\rho\oeq\O(\rho^2),
\ee
and the corresponding symplectic current densities on the conformal boundary are therefore
\be\label{FG symplectic currents}
\omega^\rho=\lim_{\rho\to0}\delta\Theta^\rho=\delta\varphi\delta\big(\ell^2\ddot{\varphi}-\varphi''\big)\oeq\ell^2\partial_u\big(\delta\varphi\delta\dot{\varphi}\big),
\q\q
\check{\omega}^\rho=\lim_{\rho\to0}\delta\check{\Theta}^\rho=0.
\ee
As expected, we see that $\varphi$ enters $\omega^\rho$ as a simple corner contribution $\partial_t(\dots)$. By comparing this result with \eqref{BS symplectic currents}, one can see that the EH and CS symplectic currents are swapping their roles. In BS gauge the EH symplectic current vanishes on $\B$, while in FG gauge it is the CS symplectic current which vanishes. Similarly, the boundary conformal factor contributes to the CS symplectic current in BS gauge and to the EH symplectic current in FG gauge. It may come as a surprise that the symplectic current on $\B$ depends so drastically on the choice of coordinates. We will however verify and confirm this directly in section \ref{sec:diffeo} using the field-dependent diffeomorphism between the two gauges.

\subsection{Residual diffeomorphisms}

We now look for the asymptotic Killing vectors $\xi^\mu=(\xi^\rho,\xi^a)$ which preserve the FG gauge conditions \eqref{FG gauge conditions}. We find
\bsub\label{FG AKV gauge conditions}
\be
\L_\xi g_{\rho\rho}=0\q&\Rightarrow\q\xi^\rho=-h\rho,\\
\L_\xi g_{\rho a}=0\q&\Rightarrow\q\xi^a=Y^a+\ell^2\partial_bh\int_0^\rho\f{\de\tilde{\rho}}{\tilde{\rho}}\gamma^{ab}(\tilde{\rho},x)=Y^a+\f{\ell^2}{2}g^{ab}_{(0)}\partial_bh\,\rho^2+\O(\rho^4),\label{FG AKV gauge conditions Ya}
\ee
\esub
where $Y^a(x^b)$ and $h(x^a)$ are the three symmetry generators. In order to bring this closer to the BS gauge, it is convenient to rename $(Y^t,Y^\phi)=(f,g)$. In addition, we also have to preserve the conformal gauge \eqref{conformal gauge}. This translates into chirality conditions for the free functions $Y^a(x^b)$. More precisely, in terms of $(f,g)$ we find the conditions
\bsub\label{FG conditions on f and g}
\be
\L_\xi g_{t \phi}^{(0)}=0\q&\Rightarrow\q\dot{g}=\f{1}{\ell^2}f',\\
\ell^2\L_\xi g_{tt}^{(0)}+\L_\xi g_{\phi\phi}^{(0)}=0\q&\Rightarrow\q\dot{f}=g',
\ee
\esub
which matches the conditions obtained in BS gauge. We now consider the same field-dependent redefinition as in \eqref{BS Weyl slicing}, i.e.
\be\label{FG Weyl slicing}
h=w-f\dot{\varphi}-g\varphi'-g',
\ee
where the new free function is now $w(x^a)$. This will guarantee later on that the Weyl charge coming from the EH Lagrangian is integrable.

As expected, we then find that the data of the solution space transforms under the residual diffeomorphisms in the same way as in BS gauge, namely
\bsub\label{FG transformation laws}
\be
\fL_\xi M&=f\dot{M}+gM'+2g'M+\f{2}{\ell^2}f'N-g''',\\
\fL_\xi N&=f\dot{N}+gN'+2g'N+2f'M-f''',\\
\fL_\xi\varphi&=w.\label{FG transformation law varphi}
\ee
\esub
The algebra of the vector fields is obtained once again from the modified bracket
\be
\big[\xi_1,\xi_2\big]_*=\big[\xi_1,\xi_2\big]-\fL_{\xi_1}\xi_2+\fL_{\xi_2}\xi_1=\xi(f_{12},g_{12},w_{12}),
\ee
and in the case where $\fL_\xi f=0$, $\fL_\xi g=0$, and $\fL_\xi w=0$, we find
\bsub
\be
f_{12}&=f_1g'_2+g_1f'_2-(1\leftrightarrow2),\\
g_{12}&=\f{1}{\ell^2}f_1f'_2+g_1g'_2-(1\leftrightarrow2),\\
w_{12}&=0.
\ee
\esub
The upshot of this construction is that we have written the solution space and the symmetry generators in FG gauge in a manner which matches exactly the notations and the equations of the BS gauge, apart from the form of the metric itself.

\subsection{Charges}

In order to compute the charges, we use once again the expressions given in \eqref{charge aspects}, but this time with $\mu,\nu=\rho,t$. The charge aspects deriving from the EH and metric CS Lagrangians are both found to be integrable and given by
\be\label{FG charges}
Q_\xi=\lim_{\rho\to0}\Q_\xi^{\rho t}\oeq fM+gN+\ell^2(\dot{w}\varphi-w\dot{\varphi}),
\q\q
\check{Q}_\xi=\lim_{\rho\to0}\check{\Q}_\xi^{\rho t}\oeq\f{1}{\ell^2}fN+gM.
\ee
We can notice in particular that
\be
\check{Q}_\xi=Q_\xi\left(f\to g\,,\,g\to\f{f}{\ell^2}\,,\,w\to0\right),
\ee
which is in a sense dual to the relation \eqref{BS duality} obtained in BS gauge. Note that the integrability of the EH charge is made possible by the change of slicing \eqref{FG Weyl slicing} for the Weyl transformations.

As announced, one can see that in FG gauge it is the EH Lagrangian which gives rise to a non-vanishing Weyl charge, while the metric CS Lagrangian has no Weyl charge. This is precisely opposite to what happens in BS gauge. Just like in BS gauge, one can show that the Weyl contribution to the charges arises from a corner symplectic potential. Using $\vartheta=\ell^2\dot{\varphi}\delta\varphi$ we find
\be
I_\xi(\delta\vartheta)=\ell^2I_\xi(\delta\dot{\varphi}\delta\varphi)=\ell^2\delta(\dot{w}\varphi-w\dot{\varphi}),
\ee
which is the Weyl contribution to $Q_\xi$ in \eqref{FG charges}.

Using the evolution equations \eqref{FG evolution of M and N} and the conditions \eqref{FG conditions on f and g}, we find that the flux coming from the charges \eqref{FG charges} is
\be
\dot{Q}_\xi\oeq\ell^2(\ddot{w}\varphi-w\ddot{\varphi}),
\q\q
\dot{\check{Q}}_\xi\oeq0.
\ee
Consistently, one can check that this time evolution is the one derived from the symplectic flux. Indeed, using \eqref{FG symplectic currents} and \eqref{FG transformation laws} we find\footnote{Here, at the difference with \eqref{BS symplectic flux}, a minus sign is required in order to relate the symplectic flux to the time evolution of the charges, because the charges are obtained in \eqref{FG charges} from $\slashed{\delta}\Q^{\rho t}_\xi$ instead of $\slashed{\delta}\Q^{t\rho}_\xi=-\slashed{\delta}\Q^{\rho t}_\xi$.}
\be
\delta\dot{Q}_\xi\oeq-I_\xi\omega^\rho,
\q\q
\delta\dot{\check{Q}}_\xi\oeq-I_\xi\check{\omega}^\rho.
\ee
Finally, the algebra of these charges in FG gauge is found to be
\bsub
\be
\lb Q_{\xi_1},Q_{\xi_2}\rb=\fL_{\xi_2}Q_{\xi_1}&\oeq Q_{[\xi_1,\xi_2]_*}-\big(f_1g_2'''+\ell^2w_1\dot{w}_2\big)-(1\leftrightarrow2),\\
\lb\check{Q}_{\xi_1},\check{Q}_{\xi_2}\rb=\fL_{\xi_2}\check{Q}_{\xi_1}&\oeq\check{Q}_{[\xi_1,\xi_2]_*}-\f{1}{2}\left(\f{1}{\ell^2}f_1f_2'''+g_1g_2'''\right)-(1\leftrightarrow2),
\ee
\esub
where at the difference with \eqref{BS charge algebra} the Weyl parameter is now appearing in the Heisenberg central extension of the EH charges.

The comparison of the BS and FG gauges in (A)dS$_3$, and in particular of their symplectic structure and charges, perfectly exemplifies the leitmotiv of this work. Indeed, we have found that two gauges with the same asymptotic fall-offs produce different asymptotic charges. This difference relies on the presence or absence of the Weyl charge, while the charges associated to the mass and angular momentum are untouched. We proposed that the Weyl charge is an instance of a ``kinematical" charge, which has thus a different status from the other charges. In the next section, we will show that it is indeed arising from the diffeomorphism mapping between these two gauges, which is henceforth a \textit{large} diffeomorphism.

\section{Diffeomorphism between Bondi--Sachs and Fefferman--Graham}
\label{sec:diffeo}

In the previous two sections we have studied the solution spaces for (A)dS$_3$ gravity with boundary Weyl freedom in BS and FG gauge. In each gauge, we have done this study by starting from the explicit form of the metric and then computing various quantities of interest (such as the symplectic current and the charges). Alternatively, it is also possible to use the explicit diffeomorphism between the two gauges in order to relate these various quantities directly. This construction is however subtle because of the field-dependent nature of the diffeomorphism. We illustrate this below with the simplest example where the subtlety appears, which is when mapping the asymptotic Killing vector between the two gauges. The transformation of the vector field is non-tensorial and given by the modified formula \eqref{change of AKV compact}, where the second term arises from the field-dependency of the diffeomorphism. Without taking this extra term into account, one would (wrongly) conclude that the asymptotic symmetries in FG gauge have no Weyl freedom. This illustrates once again how the field-dependency of the diffeomorphism is crucial and must be taken into account to properly map between the solution spaces. We also give in \eqref{transformation theta} the non-tensorial transformation formula for the EH pre-symplectic potential, and verify explicitly that it reproduces the results \eqref{theta BS} and \eqref{theta FG} computed above.

Importantly, the diffeomorphism between the BS and FG gauge can be used to map the BS gauge with $\varphi=0$ to the FG gauge with $\varphi\neq0$. This is because the diffeomorphism naturally includes a Weyl rescaling of the boundary metric. Since we have shown above (we focus here on the EH Lagrangian) that the boundary Weyl charge is non-vanishing in FG gauge, while it always vanishes in BS gauge (even if the Weyl freedom is included in the solution space), one can conclude that a field-dependent diffeomorphism between two gauges can breathe life into surface charges. This is precisely what is exemplified by the present calculation.

\subsection{Exact diffeomorphism between the zero modes}

Let us start by considering the diffeomorphism between the zero mode solutions in BS and FG gauge, which are obtained by taking the fields $(M,N,\varphi)$ as constants. In particular, the zero mode of the solution \eqref{BS-Weyl line element} is
\be
\de s^2=-\left(e^{2\varphi}\f{r^2}{\ell^2}-2M\right)\de u^2-2e^\varphi\de u\,\de r+2N\,\de u\,\de\theta+e^{2\varphi}r^2\de\theta^2.
\ee
Using the diffeomorphism
\be\label{finite diffeo1}
u=\ell\phi+F(\rho),
\q\q
r=R(\rho),
\q\q
\theta=\f{t}{\ell}+T(\rho),
\ee
with
\bsub\label{finite diffeo2}
\be
F(\rho)&=f_0-\f{e^{2\varphi}}{N}\int_1^\rho R(p)^2\partial_pT(p)\,\de p,\\
R(\rho)&=\f{e^{\varphi_0-\varphi}}{2\rho}\sqrt{\displaystyle 4e^{-2\varphi_0}M\ell^2\rho^2-e^{-4\varphi_0}\Delta^2\ell^2\rho^4-4}\,,\\
T(\rho)&=t_0+\f{\ell N}{2\Delta}\left(\f{\displaystyle\text{arctanh}\left(\f{e^\varphi R(\rho)}{\sqrt{\ell(\ell M-\Delta)}}\right)}{\sqrt{\ell(\ell M-\Delta)}}-\f{\displaystyle\text{arctanh}\left(\f{e^\varphi R(\rho)}{\sqrt{\ell(\ell M+\Delta)}}\right)}{\sqrt{\ell(\ell M+\Delta)}}\right),
\ee
\esub
and $\Delta=\sqrt{\ell^2M^2-N^2}$, we end up with the line element
\be
\de s^2=\ell^2\rho^{-2}\de\rho^2+\gamma_{ab}\,\de x^a\,\de x^b,
\ee
where
\be
\gamma_{ab}=
\begin{pmatrix}
\displaystyle-\f{e^{2\varphi_0}}{\ell^2}\rho^{-2}+M-\f{1}{4}e^{-2\varphi_0}\Delta^2\rho^2&N\\
N&\displaystyle e^{2\varphi_0}\rho^{-2}+\ell^2M+\f{1}{4}e^{-2\varphi_0}\ell^2\Delta^2\rho^2
\end{pmatrix}.
\ee
This solution in FG gauge is the zero mode of the solution built in section \ref{sec:FG solution} above, with the identification $\varphi=\varphi_0$.

One can see that the diffeomorphism \eqref{finite diffeo1}-\eqref{finite diffeo2}, although it maps between two relatively elementary exact solutions of three-dimensional gravity, is rather ``complicated''. It may seem puzzling that \eqref{finite diffeo2} sends $u$ to $\phi$ and $\theta$ to $t$ (i.e. that it swaps time and angle when switching from BS to FG gauge). It is indeed possible to write another diffeomorphism which maps between the zero modes while sending $u$ to $t$ and $\theta$ to $\phi$, but its expression involves elliptic integrals and is too lengthy to be displayed. In any case, regardless of the form of the diffeomorphism which is used, the important thing to notice is that it gives the possibility of performing a Weyl rescaling while changing the gauge. This freedom is encoded in the function $\varphi_0$. Using this freedom, it is possible to reach the FG gauge with Weyl conformal factor even when starting from the BS gauge without Weyl factor. As we will now show, this freedom also exists when constructing the diffeomorphism between the full solution spaces where $(M,N,\varphi)$ are local fields. 

\subsection{Perturbative diffeomorphism between the solution spaces}

Our goal is now to perform the diffeomorphism from the general solution \eqref{BS-Weyl line element} in BS gauge to the general solution \eqref{general FG metric} in FG gauge. We will first study the diffeomorphism between the line elements, and then its action on the residual diffeomorphisms.

\subsubsection{Transforming the metric}

To map the general solution \eqref{BS-Weyl line element} from BS to FG gauge we proceed as in \cite{Poole:2018koa,Compere:2019bua,Ciambelli:2020ftk}. First, we go from BS coordinates $(u,r,\theta)$ to the tortoise coordinates $(t^*,r^*,\phi^*)$ defined by
\be\label{tortoise coordinates}
u\coloneqq t^*-r^*,
\q\q
r\coloneqq-\ell\cot\left(\f{r^*}{\ell}\right),
\q\q
\theta\coloneqq\phi^*.
\ee
The boundary at $r=\infty$ is then mapped to $r_*=0$. We then expand the resulting line element in tortoise coordinates around $r_*=0$, and finally go to FG gauge \eqref{FG gauge conditions} perturbatively using the diffeomorphism
\be\label{tortoise to FG diffeo}
t^*=t+\sum_{n=1}^\infty\rho^nT_n(x^a),
\q\q
r^*=\sum_{n=1}^\infty\rho^nR_n(x^a),
\q\q
\phi^*=\phi+\sum_{n=1}^\infty\rho^nF_n(x^a).
\ee
Then, by fixing the three functions $(T_n,R_n,F_n,)$ in this expansion at order $\rho^n$, one can tune terms at order $\rho^{n-3}$ in the metric in order to satisfy the three FG gauge conditions \eqref{FG gauge conditions}. Explicitly, for $n=1$ we first find
\bsub
\be
g_{\rho\phi}&=\O(\rho^{-1})\hspace{-2cm}&&\Rightarrow\q F_1=0,\\
g_{\rho t}&=\O(\rho^{-1})\hspace{-2cm}&&\Rightarrow\q T_1=R_1(1-e^{-\varphi}),\\
g_{\rho\rho}&=\ell^2\rho^{-2}+\O(\rho^{-1}),\hspace{-2cm}&&
\ee
\esub
where the last condition follows automatically from the previous two. Importantly, this leaves $R_1$ free, which can therefore be redefined as $R_1=\ell^2e^{-\varphi_0}$ in terms of a free function $\varphi_0(x^a)$. We then obtain
\be
\de s^2=\ell^2\rho^{-2}\de\rho^2+\gamma_{ab}\,\de x^a\,\de x^b+\O(\rho^{-1}),
\ee
with
\be
\gamma_{ab}=\rho^{-2}g_{ab}^{(0)},
\q\q
g_{ab}^{(0)}=
e^{2(\varphi+\varphi_0)}
\begin{pmatrix}
-1/\ell^2&0\\
0&1
\end{pmatrix},
\ee
which shows that $\varphi_0$ is the Weyl freedom included in the diffeomorphism from BS to FG gauge. Since we have started from the BS gauge with $\varphi\neq0$, we will set $\varphi_0=0$ for what follows, but one should keep in mind that this free function can be turned on in order to reach an FG gauge with Weyl factor when starting from a BS gauge with $\varphi=0$. In other words, the FG metric with Weyl factor $\varphi$ can be obtained in two equivalent ways:\\
1) starting from a BS metric with $\varphi\neq0$ and performing a diffeomorphism with $\varphi_0=0$,\\
2) starting from a BS metric with $\varphi=0$ and performing a diffeomorphism with $\varphi_0=\varphi$.\\
It is precisely this last possibility which makes it explicit that the diffeomorphism with $\varphi_0=\varphi$ is large and turns on the Weyl charge.

If we then continue with the construction of the perturbative diffeomorphism, we find for $n=2$ the conditions
\bsub
\be
g_{\rho\phi}&=\O(\rho^0)\hspace{-2cm}&&\Rightarrow\q F_2=\f{\ell^2}{2}\varphi'e^{-2\varphi},\\
g_{\rho t}&=\O(\rho^0)\hspace{-2cm}&&\Rightarrow\q T_2=-\f{\ell^4}{2}\dot{\varphi}e^{-2\varphi}(1+2e^\varphi),\\
g_{\rho\rho}&=\ell^2\rho^{-2}+\O(\rho^0),\hspace{-2cm}&&\Rightarrow\q R_2=-\ell^4\dot{\varphi}e^{-\varphi}.
\ee
\esub
More generally, by solving recursively one can get
\bsub
\be
g_{\rho\phi}&=\O(\rho^n)\hspace{-2cm}&&\Rightarrow\q F_{i\leq n+2}\,=(\dots),\\
g_{\rho t}&=\O(\rho^n)\hspace{-2cm}&&\Rightarrow\q T_{i\leq n+2}\,=(\dots),\\
g_{\rho\rho}&=\ell^2\rho^{-2}+\O(\rho^n),\hspace{-2cm}&&\Rightarrow\q R_{i\leq n+2}=(\dots).
\ee
\esub
Solving up to $n=3$, we finally end up with
\be
\de s^2=\ell^2\rho^{-2}\de\rho^2+\gamma_{ab}\,\de x^a\,\de x^b+\O(\rho^3),
\q\q
\gamma_{ab}=\rho^{-2}g_{ab}^{(0)}+g_{ab}^{(2)}+\rho g_{ab}^{(3)}+\rho^2g_{ab}^{(4)}.
\ee
Requiring $g_{ab}^{(3)}=0$ then imposes the equations of motion $\eqref{BS evolution of M and N}$. Indeed, the condition $g_{t\phi}^{(3)}=0$ requires to impose $\dot{N}=M'$, while $g_{\phi\phi}^{(3)}=0$ requires $\dot{M}=N'/\ell^2$. After imposing this, we finally obtain
\be\label{BStoFG}
\de s^2_\text{BS}=\de s^2_\text{FG}+\O(\rho^3),
\ee
and the subleading terms in $\rho$ can be made vanishing by tuning $(R_n,T_n,F_n)_{n\geq6}$. At the end of the day, the coordinate change from BS to FG gauge has the expansion
\bsub\label{xBS as xFG}
\be
u&=t-\ell^2e^{-\varphi}\rho+\O(\rho^2),\\
r&=-\f{1}{\rho}-\ell^2e^{-\varphi}\dot{\varphi}+\f{\ell^2}{4}e^{-2\varphi}\rho\Big((\varphi')^2-2M+\ell^2\big(2\ddot{\varphi}-3\dot{\varphi}^2\big)\Big)+\O(\rho^2),\\
\theta&=\phi+\O(\rho^2).
\ee
\esub
Even if the field dependency enters only at subleading order, as we have seen with the computations above this diffeomorphism can turn on a codimension-two charge.

\subsubsection{Transforming the asymptotic Killing vector}
\label{sec:diffeo vector}

Let us now study how the above diffeomorphism can be used to map the asymptotic Killing vector \eqref{BS vector field} from the BS gauge $\xi^\mu_\text{BS}=(\xi^u,\xi^r,\xi^\theta)$ to the FG gauge $\xi^\mu_\text{FG}=(\xi^t,\xi^\phi,\xi^\rho)$. Again, we proceed by first going from the BS coordinates to the tortoise coordinates \eqref{tortoise coordinates}, and then expand around $r^*=0$ before moving to FG coordinates using \eqref{tortoise to FG diffeo}.

Since the change of coordinates \eqref{tortoise to FG diffeo} is field-dependent, the transformation of the vector field has to be computed using the adjusted formula \cite{Compere:2016hzt} (see equation (36) there)
\be\label{change of AKV}
\xi^\mu_\text{FG}(x_{\text{FG}})=\f{\partial x^\mu_\text{FG}(x_{\text{BS}})}{\partial x^\nu_\text{BS}}\big(x_{\text{BS}}(x_{\text{FG}})\big)\big(\xi^\nu_\text{BS}(x_{\text{BS}}(x_{\text{FG}}))+\fL_\xi x^\nu_\text{BS}(x_{\text{FG}})\big).
\ee
This formula is written here in FG coordinates, and we have made completely explicit the coordinate dependency of the various terms. In particular, the vector field $\xi^\mu_\text{BS}$ is given by \eqref{BS vector field} in BS coordinates, and must be written in FG coordinates using $x^\mu_\text{BS}(x_\text{FG})$ given by \eqref{xBS as xFG}. In formula \eqref{change of AKV}, the first term is the ``standard'' diffeomorphism transformation of $\xi^\mu_\text{BS}$ given by the action of the Jacobian, and the second term is here because of the field-dependency of the vector field. This correction term is non-vanishing because it is computed as the action of $\fL_\xi$ in \eqref{FG transformation laws} on the field-dependency which appears in the right-hand side of \eqref{xBS as xFG}.

For compactness and for later convenience, let us denote the initial coordinates by $\tilde{x}^\mu$ and the new ones after the diffeomorphism by $x^\mu$. In the present case we have $\tilde{x}^\mu=x^\mu_\text{BS}$ and $x^\mu=x^\mu_\text{FG}$. Let us then denote the Jacobian and its inverse by
\be
\Ju^\mu_\alpha\coloneqq\f{\partial\t{x}^\mu}{\partial x^\alpha}=\partial_\alpha\t{x}^\mu,
\qquad
\Jd^\mu_\alpha\coloneqq\f{\partial x^\mu}{\partial\t{x}^\alpha}=\t{\partial}_\alpha x^\mu=\big(\Ju^{-1}\big)^\mu_\alpha.
\ee
With this notation, formula \eqref{change of AKV} takes the form
\be\label{change of AKV compact}
\xi^\mu=\zeta^\mu+\Jd^\mu_\alpha\fL_\xi\tilde{x}^\alpha,
\q\q
\zeta^\mu\coloneqq\Jd^\mu_\alpha\tilde{\xi}^\alpha,
\ee
where $\zeta^\mu$ is the ``standard'' piece obtained in the field-independent case. It is instructive to look at the two contributions separately. The first term is
\bsub
\be
\zeta^t&=f-\ell^2e^{-\varphi}\big(h+f\dot{\varphi}+g\varphi'+\dot{f}\big)\rho+\O(\rho^2),\\
\zeta^\phi&=g-\ell^2e^{-\varphi}\big(\dot{g}-\ell^{-2}f'\big)\rho+\O(\rho^2),\\
\zeta^\rho&=-h\rho+\O(\rho^2),
\ee
\esub
where $(f,g,h)$ are functions of $x^a=x^a_\text{FG}=(t,\phi)$. If we were to forget about the second term in \eqref{change of AKV compact} arising from the field-dependency, we would interpret $\zeta^\mu$ as the residual vector field in FG gauge. If this were to be the case, the vector field would have to preserve the gauge condition $g_{\rho a}=0$, which would require to impose $h\stackrel{!}{=}-(f\dot{\varphi}+g\varphi'+\dot{f})$. We have indeed shown in \eqref{FG AKV gauge conditions Ya} that there is no term of order $\rho$ in the FG residual vector field. However, imposing this condition is equivalent to setting $w\stackrel{!}{=}0$ according to \eqref{FG Weyl slicing}. This would in turn wrongly imply that there is no symmetry parameter for the Weyl transformations in FG gauge, and therefore that the Weyl component of the charge is vanishing! Turning this argument around, one clearly sees that the field dependency of the diffeomorphism is responsible for ``turning on'' the Weyl charge when going from BS to FG gauge.

In order to obtain the correct result for the asymptotic Killing vector in FG gauge, let us now take into account the contribution of the second term in \eqref{change of AKV compact}. For this, we compute
\be
\Jd^\mu_\alpha=
\begin{pmatrix}
1-\ell^2e^{-\varphi}\dot{\varphi}\rho&0&-\ell^2e^{-\varphi}\varphi'\rho\\
0&0&1\\
0&0&0
\end{pmatrix}+\O(\rho^2),
\q\quad
\fL_\xi\tilde{x}^\alpha=
\begin{pmatrix}
\ell^2e^{-\varphi}\fL_\xi\varphi\rho\\
\O(1)+\O(\rho)\\
0
\end{pmatrix}+\O(\rho^2),
\ee
where the second component of $\fL_\xi\tilde{x}^\alpha$ is lengthy but irrelevant for the present calculation. Putting this together and using \eqref{FG transformation law varphi} with \eqref{FG Weyl slicing}, we find that
\be
\Jd^\mu_\alpha\fL_\xi\tilde{x}^\alpha=
\begin{pmatrix}
\ell^2e^{-\varphi}\big(h+f\dot{\varphi}+g\varphi'+g'\big)\rho\\
0\\
0
\end{pmatrix}+\O(\rho^2),
\ee
which can then be added to $\zeta^\mu$ to finally obtain the FG residual vector field
\bsub
\be
\xi^t&=f-\ell^2e^{-\varphi}\big(\dot{f}-g'\big)\rho+\O(\rho^2),\\
\xi^\phi&=g-\ell^2e^{-\varphi}\big(\dot{g}-\ell^{-2}f'\big)\rho+\O(\rho^2),\\
\xi^\rho&=-h\rho+\O(\rho^2).
\ee
\esub
Consistently, one then finds that the vanishing of the terms of order $\rho$ in $\xi^a$ imply the conditions \eqref{FG conditions on f and g} (alternatively one may also impose these conditions from the onset since they are already true for the initial vector field in BS gauge). By performing this computation to sufficiently low order, one can explicitly verify that it reproduces the expected result \eqref{FG AKV gauge conditions} for the residual vector field in FG gauge. This result already hints at the deeper interplay between the field dependency of the diffeomorphism and the transformation laws of objects of the covariant phase space.

\subsubsection{Transforming the symplectic potential}

With the example of the vector field in mind, we can now use \eqref{change of AKV compact} in order to compute the transformation of more involved objects, such as the EH symplectic potential. This will enable us to illustrate once again the non-tensorial transformation behavior due to the field-dependency of the diffeomorphism, and to recover the two results \eqref{theta BS} and \eqref{theta FG} in BS and FG gauge.

Using the same notations as above, we start from a gauge in which the coordinates are denoted $\tilde{x}^\mu$. In these coordinates the EH pre-symplectic density is given by\footnote{We drop the subscript EH from now on.}
\be\label{tilde Theta}
\tilde{\Theta}^\mu=\f{1}{2}\sqrt{-\tilde{g}}\,\Big(\tilde{g}^{\alpha\beta}\delta\tilde{\Gamma}^\mu_{\alpha\beta}-\tilde{g}^{\mu\alpha}\delta\tilde{\Gamma}^\beta_{\alpha\beta}\Big).
\ee
We then have the formulas
\be
\sqrt{-\tilde{g}}&=|\Ju|^{-1}\sqrt{-g},\\
\tilde{g}^{\alpha\beta}&=\Ju^\alpha_\mu\Ju^\beta_\nu g^{\mu\nu},\\
\tilde{\Gamma}^\mu_{\alpha\beta}&=\Jd^\rho_\alpha\Jd^\sigma_\beta\big(\Ju^\mu_\lambda\Gamma^\lambda_{\rho\sigma}-\partial_\rho\Ju^\mu_\sigma\big),\\
\tilde{\Gamma}^\beta_{\alpha\beta}&=\Jd^\rho_\alpha\big(\Gamma^\sigma_{\rho\sigma}-\Jd^\sigma_\beta\partial_\rho\Ju^\beta_\sigma\big).
\ee
Now, the crucial observation is that the Jacobian has a non-vanishing field-space variation because of the field-dependency of the diffeomorphism. More precisely, we have
\be\label{deltaJ}
\delta\Ju^\mu_\alpha&=\delta\left(\f{\partial\t{x}^\mu}{\partial x^\alpha}\right)=\f{\partial\delta\t{x}^\mu}{\partial x^\alpha}=\partial_\alpha\delta\t{x}^\mu,
\ee
where we have however assumed that $[\partial_\alpha,\delta]=0$, meaning that the coordinates in the new gauge are field-independent. This is the same logic as when performing a change of slicing as in \eqref{BS Weyl slicing}, where we use a field-dependent redefinition of the initial symmetry parameter and then assume that the new symmetry parameter ($w$ in the present case) is field-independent. It is precisely the non-vanishing of \eqref{deltaJ} which implies that $(n\geq1)$-forms in field-space transform as connections rather than tensors. This transformation behavior is fundamental in order to ensure that the phase space is correctly mapped between the two gauges.

Let us now illustrate this by deriving the transformation behavior of the pre-symplectic potential. Starting from \eqref{tilde Theta} and using the above formulas, we get
\be
%%%%%%%%%
\tilde{\Theta}^\mu
%%%%%%%%%
&=\f{1}{2}\sqrt{-\tilde{g}}\,\Big(\tilde{g}^{\alpha\beta}\delta\tilde{\Gamma}^\mu_{\alpha\beta}-\tilde{g}^{\mu\alpha}\delta\tilde{\Gamma}^\beta_{\alpha\beta}\Big)\cr
%%%%%%%%%
&=\f{1}{2}|\Ju|^{-1}\sqrt{-g}\,\Ju^\alpha_\tau\Big(\Ju^\beta_\nu g^{\tau\nu}\delta\Big[\Jd^\rho_\alpha\Jd^\sigma_\beta\big(\Ju^\mu_\lambda\Gamma^\lambda_{\rho\sigma}-\partial_\rho\Ju^\mu_\sigma\big)\Big]-\Ju^\mu_\nu g^{\tau\nu}\delta\Big[\Jd^\rho_\alpha\big(\Gamma^\sigma_{\rho\sigma}-\Jd^\sigma_\beta\partial_\rho\Ju^\beta_\sigma\big)\Big]\Big)\cr
%%%%%%%%%
&=|\Ju|^{-1}\Ju^\mu_\alpha\Theta^\alpha+|\Ju|^{-1}\A^\mu,
%%%%%%%%%
\ee
where
\be\label{A anomaly}
\!\!\A^\mu=\f{1}{2}\sqrt{-g}\,g^{\tau\nu}\Ju^\alpha_\tau\Big(\Ju^\beta_\nu\Big\{\Gamma^\lambda_{\rho\sigma}\delta\Big[\Jd^\rho_\alpha\Jd^\sigma_\beta\Ju^\mu_\lambda\Big]-\delta\Big[\Jd^\rho_\alpha\Jd^\sigma_\beta\partial_\rho\Ju^\mu_\sigma\Big]\Big\}+\Ju^\mu_\nu \Big\{\delta\Big[\Jd^\rho_\alpha\Jd^\sigma_\beta\partial_\rho\Ju^\beta_\sigma\Big]-\Gamma^\sigma_{\rho\sigma}\delta\Jd^\rho_\alpha\Big\}\Big).
\ee
Note that this correction term is arising solely from the variations of the coordinates. Given that in \eqref{xBS as xFG} we have $\tilde{x}_\text{BS}(x_\text{FG})$ instead of $x_\text{FG}(\tilde x_\text{BS})$, it is more convenient to write the transformation of the potential the other way around, i.e. as
\be\label{transformation theta}
\Theta^\mu_\text{FG}=|\Ju|\Jd^\mu_\alpha\tilde{\Theta}^\alpha_\text{BS}-\Jd^\mu_\alpha\A^\alpha,
\ee
so that the right-hand side can be evaluated in FG coordinates using $\tilde{x}_\text{BS}(x_\text{FG})$. As announced, this formula reveals the non-tensorial transformation behavior of the pre-symplectic potential due to the field-dependency of the diffeomorphism. In fact, there is an extra subtlety to be addressed before being able to apply this formula in order to check how $\Theta^r$ in \eqref{theta BS} gets mapped to $\Theta^\rho$ in \eqref{theta FG}. This subtlety arises because we need to evaluate the right-hand side of \eqref{transformation theta} in FG coordinates using the field-dependent change of coordinates $\tilde{x}_\text{BS}(x_\text{FG})$. This implies that in the computation of $\tilde{\Theta}_\text{BS}$ one must keep track of the variations $\delta\tilde{x}_\text{BS}$ of the initial coordinates, so that the field-dependent relation $\tilde{x}_\text{BS}(x_\text{FG})$ can then be implemented consistently. This also means that when computing $\tilde{\Theta}_\text{BS}$ one must recall that $[\tilde{\partial}_\alpha,\delta]\neq0$. This is a crucial point which, as we will comment upon in the next section, challenges foundational assumptions at the heart of the variational bi-complex framework. Finally, note that the very fact that we can turn on a charge via a diffeomorphism between gauges implicitly already implies that the symplectic potential \textit{must} transform non-tensorially.

With this in mind, we can now compute the terms in \eqref{transformation theta}. In order to avoid lengthy expressions, let us do so up to $\O(1)$ terms, since it already illustrates the mechanism. First, from \eqref{theta FG}, which is the direct computation in FG gauge, we get that
\be\label{FG potential O(1)}
\Theta^\rho_\text{FG}\oeq\f{1}{\ell^2\rho^2}\delta e^{2\varphi}+\O(1).
\ee
Using the ingredients above, we also get
\bsub
\be
|\Ju|\Jd^\mu_\alpha\tilde{\Theta}^\alpha_\text{BS}&=\f{1}{\ell^2\rho^2}\delta e^{2\varphi}+\f{1}{2\rho}e^\varphi\big(3\dot{\varphi}\delta\varphi-\delta\dot{\varphi}\big)+\O(1),\\
\Jd^\mu_\alpha\A^\alpha&=\f{1}{2\rho}e^\varphi\big(3\dot{\varphi}\delta\varphi-\delta\dot{\varphi}\big)+\O(1),
\ee
\esub
which when subtracted gives \eqref{FG potential O(1)} as announced. One can check that this yields the correct result to all orders in $\rho$ and for all the components of the potential. Interestingly, one can see that the tensorial piece of the transformation contains a terms in $\rho^{-1}$, which is precisely canceled by the field-dependent contribution. This piece can (in part) be traced back to the order $r$ term in the BS potential \eqref{theta BS}. This example clearly illustrates the role of the field-dependent transformation rules in order to map the phase spaces and variational quantities between the two gauges. 

In summary, we have demonstrated how to map the BS gauge and its phase space to the FG gauge and its phase space. We showed that the diffeomorphism relating these gauges is a large diffeomorphism, as it activates the Weyl charge in the FG gauge. This finding highlights the kinematical nature of the Weyl charge, in contrast with the dynamical charges associated with mass and angular momentum, which transform covariantly when changing the gauge. Furthermore, we have analyzed the field dependency present in the diffeomorphism mapping between the two gauges. Specifically, field-space zero-forms, such as the metric, transform covariantly, while residual symmetries and higher forms transform as connections. This distinction arises because the BS coordinates, when expressed as functions of FG coordinates, do not commute with field variations. While a more systematic investigation of this phenomenon and its implications for the theory of asymptotic symmetries is left for future work, the Weyl charge example discussed here provides an initial insight into this direction, opening avenues for further explorations.

\section{Conclusions}
\label{sec:conclusion}

In this work, we have examined an example which highlights a remarkable property of codimension-two charges: under specific conditions, these charges may emerge or vanish when transitioning between different gauges. As far as we are aware, the simplest setup in which this happens is the one presented here, namely (A)dS$_3$ gravity with boundary conditions allowing for Weyl rescalings, where the Weyl charge is vanishing in BS gauge \eqref{BS line element} but non-vanishing in FG gauge \eqref{general FG metric} when working in EH gravity. When starting from the metric CS Lagrangian \eqref{EH and CS Lagrangians} this situation is reversed. Two other properties can be used to characterize these Weyl charges. The first one is the absence of flux-balance law from the Einstein field equations. This is in sharp contrast with the mass and angular momentum, which satisfy the evolution equations \eqref{BS evolution of M and N}. The second one is the fact that the Weyl charge can be seen as arising from a corner term in the symplectic structure. For these reasons, we have suggested to label the Weyl charge as ``kinematical''. In total, this charge therefore exhibits three properties: 1) it can be vanishing or not depending on the gauge; 2) it has no associated flux-balance law; 3) it can be rewritten as a corner contribution to the symplectic structure. While criteria 2) and 3) can be diagnosed \textit{within} a given gauge, criterion 1) requires to exhibit a gauge in which the charge is vanishing, and to potentially study the diffeomorphism between the gauges.

We have studied in section \ref{sec:diffeo} the diffeomorphism between the BS and FG gauges, and highlighted in particular the key role played by its field-dependency. While this does not actually explain \textit{why} the Weyl charge has the characteristics of a kinematical charge, it enables to clearly see the mechanism at play when changing the coordinate gauge and turning on or off the Weyl charge. The key feature is that the field-dependency of the diffeomorphism implies a non-tensorial transformation behavior for the asymptotic Killing vector and the field-space forms (such as the symplectic potential or the Iyer--Wald charge aspect). As we have shown in section \ref{sec:diffeo vector}, by forgetting about the field-dependency and the ensuing extra term in formula \eqref{change of AKV compact}, one would wrongly conclude that the Weyl rescaling symmetry does not carry over from the BS to the FG gauge. Instead, the field-dependency is crucial in order to properly map the residual vector fields and their symmetry parameters between the gauges. We have also shown how to perform a diffeomorphism of the pre-symplectic potential, which is found to transform as in \eqref{transformation theta}, i.e. as a vector density with the addition of a correction term due to the variations of the field-dependent change of coordinates. This gives results that are consistent with the presence or absence of symplectic flux depending on the chosen gauge (which itself is a direct consequence of the presence or absence of a kinematical charge).

Building up on the present example as a guiding case study, several interesting questions are left to be investigated for future work:
\begin{itemize}
%%%%%%%%%%%
\item[$\bullet$] \textbf{Examples of kinematical charges.} First, it will be useful to exhibit and study other examples of kinematical charges in order to corroborate the conclusions presented here concerning their properties. There are already know situations in which such charges seem to appear.

For four-dimensional asymptotically-flat spacetimes, it has been shown in \cite{Geiller:2024amx} that the so-called partial Bondi gauge \cite{Geiller:2022vto} can exhibit up to three extra surface charges (in addition to mass and angular momentum) associated to symmetries\footnote{This result holds assuming that the metric has no logarithmic terms. For polyhomogeneous solutions \cite{Geiller:2024ryw}, relaxing the Bondi--Sachs gauge to a partial Bondi gauge can also unleash new free functions in the solution space and new associated symmetries and kinematical charges (see section 3.1 of \cite{Geiller:2024amx}).}. The functions labelling these charges are $(\sqrt{q},C,D)$, where $\sqrt{q}$ is the four-dimensional Weyl factor (the determinant of the induced celestial boundary metric), and $C$ and $D$ are respectively the traces of the terms of order $r^1$ and $r^0$ in the expansion of the transverse metric. The Weyl factor $\sqrt{q}$ appears in the BMS--Weyl construction \cite{Barnich:2010eb,Freidel:2021fxf}. In principle $\sqrt{q}$ can have an arbitrary time-dependency which is not constrained by the Einstein equations, thereby satisfying one of the criteria for being related to a kinematical charge. While it is customary to work with the so-called Bondi condition $\partial_uq_{ab}=0$, which sets in turn $\partial_u\sqrt{q}=0$, this should be seen as a condition put by hand rather than a genuine flux-balance law. The trace mode $C$, on the other hand, is known to arise naturally in the so-called Newman--Unti gauge \cite{Newman:1962cia,Barnich:2011ty,Barnich:2012nkq,Geiller:2022vto}, with a corresponding symmetry parameter associated to the change of origin for the affine parameter on the null geodesics $\partial_r$. Once again, $C$ and its associated symmetry parameter appear in the asymptotic charges \cite{Geiller:2024amx}, but does not satisfy any flux-balance law. Similarly, $D$ also has an arbitrary time dependency and can appear in the charges if one relaxes the Newman--Unti gauge further. It is therefore tempting to conclude that $(\sqrt{q},C,D)$ are associated with kinematical charges. One should however investigate the fate of these charges when changing the gauge, and whether they can be obtained from corner terms in the symplectic structure.

Interestingly, the trace mode $C$ of the four-dimensional Newman--Unti gauge has a three-dimensional analogue, which was studied (and denoted $H$) in \cite{Geiller:2021vpg}. When studying the three-dimensional solution space containing both the Weyl factor and $C$, one can show that both $\varphi$ and $C$ appear in the asymptotic charges. The subtlety is that $C$ appears paired with the Weyl symmetry parameter, while $\varphi$ is paired with the Newman--Unti radial translation symmetry\footnote{This also explains why in the present work we have obtained a vanishing Weyl charge in BS gauge. The solution space described by \eqref{BS-Weyl line element} is a subspace of that studied in \cite{Geiller:2021vpg}, with $C$ and the associated radial translation turned off, so that in the asymptotic charge we see neither $C$ nor $\varphi$ (since the latter is paired with the symmetry generator of the translation of $C$).}. As one can expect, $C$ has no associated flux-balance law, and one can show that the contributions of $C$ and $\varphi$ to the charges can be removed with a choice of corner term. These are therefore examples of kinematical charges. It would be interesting to study the analogue of $C$ in the three-dimensional FG gauge.

Related examples of kinematical charges appear in generalizations of the FG gauge to accommodate Weyl transformations that are not accompanied by a transformation of the boundary coordinates. This was the rationale behind the derivation of the so-called Weyl--Fefferman--Graham (WFG) gauge in \cite{Ciambelli:2019bzz}. Later works have computed the asymptotic charges \cite{Ciambelli:2023ott,Arenas-Henriquez:2024ypo}, demonstrating that there are additional kinematical charges related to corner potentials.

We can also mention the three-dimensional derivative expansion \cite{Campoleoni:2018ltl, Ciambelli:2020ftk,Ciambelli:2020eba,Campoleoni:2022wmf}, which is a generalization of the BS gauge allowing for $g_{r\theta}\neq0$. Because of this relaxed condition, the solution space contains an extra field (the so-called boundary velocity) with an arbitrary time-dependency. It has been shown in \cite{Campoleoni:2022wmf} that the associated symmetry is a Lorentz transformation, and that there are two corner terms which can be used to turn on either this Lorentz charge or the Weyl charge. In this example there are therefore two kinematical charges a priori. Other examples of three-dimensional gauges in which kinematical charges seem to appear are \cite{Adami:2022ktn,Taghiloo:2024ewx}.

Lastly, it would be interesting to exhibit further examples of charges which can be vanishing or not depending on the Lagrangian being used. Here such an example was provided by comparing the EH and the metric CS Lagrangians in \eqref{EH and CS Lagrangians}. An immediate generalization would be to study the first order Einstein--Cartan Lagrangian. It was shown for example in \cite{Geiller:2021vpg} that the corner terms giving rise to the kinematical charges $C$ and $\varphi$ mentioned above appears naturally when working in connection-triad variables. In \cite{DePaoli:2018erh,Oliveri:2019gvm,Freidel:2020xyx,Freidel:2020svx}, it was argued that different formulations of gravity lead to symplectic structures which differ by corner terms. This implies that different Lagrangians can lead to a different charge content, although clear and simple examples of such a mechanism are lacking. Indeed, when computing for example charges associated with asymptotic BMS symmetries, both the metric Einstein--Hilbert and the connection-triad Cartan formulations lead to the same asymptotic charges. It is therefore tempting to view the kinematical charges as extra labels which could distinguish between different bulk formulations of the same theory.
%%%%%%%%%%%
\item[$\bullet$] \textbf{Criteria for dynamical versus kinematical charges.} Studying the above-mentioned examples should enable to refine the criteria which we have suggested for labeling charges as ``kinematical'' as opposed to ``dynamical''. In particular, it would be interesting to clarify whether the different properties which seem to hold for kinematical charges are actually related, and perhaps arising from a more fundamental principle. Here we have observed that the Weyl charges in (A)dS$_3$ have no flux-balance law, that they arise from corner terms (depending on the gauge), and that they can be (non)-vanishing depending on the choice of gravitational Lagrangian (here EH vs metric CS) or the choice of gauge (BS vs FG). We have also observed with the transformation laws \eqref{BS transformation laws} that dynamical charges are associated with symmetries with a global component, while kinematical charges seem to have no such global part. These properties should be investigated systematically. This should enable to obtain a clearcut classification criterion, and to study how many kinematical charges can be turned on in a given gauge. 

A particularly important open question is that of the fate of the kinematical charges in the Hamiltonian approach to asymptotic symmetries \cite{Regge:1974zd,Henneaux:2018cst,Henneaux:2018gfi,Henneaux:2018hdj,Henneaux:2019yax,Ananth:2020ojp,Fuentealba:2022xsz,Fuentealba:2023rvf,Fuentealba:2023syb}. Relatedly, one can wonder if kinematical charges at spatial infinity could play an instrumental role in connecting past to future null infinity \cite{Fiorucci:2024ndw}.

Ultimately, these questions are related to that of the \textit{physical meaning} of the kinematical charges. Since their time-dependency is arbitrary, one could \textit{choose} it by hand in such a way that it compensates the flux of the dynamical charges. This would be a way of closing a radiative system with the help of a specific boundary dynamics, very much in the spirit of edge modes \cite{Donnelly:2016auv,Geiller:2019bti,Ciambelli:2021nmv,Ciambelli:2021vnn,Carrozza:2022xut,Ball:2024hqe,Ball:2024gti,Adami:2024gdx}. Another possible use of the kinematical charges is illustrated by the Weyl charge itself, which can be used to probe anomalies of the dual boundary holographic dual \cite{Henningson:1998gx,Alessio:2020ioh,Jia:2021hgy,Jia:2023gmk}. In this sense, although the kinematical charges are not tied to the dynamics of the bulk theory as the dynamical charges (which satisfy flux-balance laws), they could still be useful auxiliary quantities allowing for spoofing certain physical properties of the system. This is a bit akin to the so-called method of image charges in electrodynamics (or differential equations), which introduces auxiliary charges and looses the boundary conditions in order to derive physical results about the initial system. Finally, one can argue that having access to a larger asymptotic symmetry group should give a better handle on quantization \cite{Freidel:2015gpa,Donnelly:2016auv,Geiller:2017xad, Speranza:2017gxd,Geiller:2017whh,Donnelly:2020xgu,Freidel:2021cjp,Ciambelli:2021vnn,Ciambelli:2021vnn,Ciambelli:2022cfr,Ciambelli:2022vot,Donnelly:2022kfs,Balasubramanian:2023dpj,Freidel:2023bnj, Ciambelli:2023bmn,Ciambelli:2024qgi}, and the kinematical charges do indeed provide representations of ``extended'' boundary symmetry groups. Ultimately, we propose that achieving a deeper understanding and systematic classification of these charges may yield more profound and wide-ranging implications than merely seeking case-by-case insights to dismiss them.
%%%%%%%%%%%
\item[$\bullet$] \textbf{Further covariantisation of the covariant phase space.} Finally, aside from the above considerations about the kinematical charges themselves, this work raises important questions on the mathematical structure of the covariant phase space, or the variational bi-complex. Indeed, we have seen that the field-dependency of the diffeomorphism from BS to FG gauge implies that transformations laws are modified with respect to the standard tensorial formulas by the addition of terms containing variations of the coordinates (or of the Jacobian). We have illustrated this with \eqref{change of AKV compact} for the vector field and with \eqref{transformation theta} for the pre-symplectic potential. Similarly, the transformation formula for the Iyer--Wald (or the Barnich--Brandt) charge aspects \eqref{charge aspects} can also be computed, but feature correction terms (in addition to the tensorial transformation) due to $\delta\Ju\neq0$ which are very intricate and lengthy. A natural question is whether there exists a way to conveniently perform such field-dependent diffeomorphisms, or to endow the covariant phase space with a structure that would enable to transform variational forms in a more geometric manner. Indeed, while the covariant phase space, as originally introduced in \cite{GawEdzki1972,Kijowski1973,Kijowski1976} and further refined in \cite{Crnkovic1986, Crnkovic1988}, does not rely on the jet bundle construction, the latter is at the core of the variational bi-complex.\footnote{The variational bi-complex has been independently constructed by Vinogradov \cite{Vinogradov1977,Vinogradov1978,Vinogradov1984,Vinogradov1984a} and Tulczyjew \cite{Tulczyjew1980}, and later improved and applied to conservation laws in \cite{MR1188434,Anderson:1996sc,Barnich:2001jy} (see also \cite{Ciambelli:2022vot} and references therein).} In the jet bundle, fields live on the fibres, while spacetime coordinates are on the base. Automorphisms of a bundle consist of changes of fibres that depend on the base point, whereas changes of base points depending on the fibres do not preserve the bundle structure. Therefore, the jet bundle is adapted to local change of fields, but not to change of coordinates depending on fields. The  diffeomorphism between our two gauges is exactly a field-dependent change of coordinates, and thus is not an automorphism of the jet bundle. Consequently, symplectic quantities do not transform as tensors under this transformation, and indeed we have seen that explicitly for the symplectic potential. To gather a more covariant construction one must therefore allow for changing of horizontal distributions in the bundle, and thus for non-trivial field-space connections.\footnote{See  \cite{Gomes:2016mwl,Gomes:2018dxs,Gomes:2018shn,Gomes:2019rgg,Gomes:2019xto} and \cite{Ciambelli:2021ujl} for works in this direction.} This requires a mathematical revisitation of the variational bi-complex, toward a more covariant manifold rather than a rigid bundle structure. This could unravel a more geometric understanding of the gauge-covariance of charges, and a more systematic treatment of the fine structure between kinematical and dynamical charges.

%%%%%%%%%%%
\end{itemize}

\subsubsection*{Acknowledgments}

We are thankful to Francesco Alessio, Glenn Barnich, Andrea Campoleoni, Arnaud Delfante, Laurent Freidel, Rob Leigh, Romain Ruzziconi, and C\'eline Zwikel for discussions. Research at Perimeter Institute is supported in part by the Government of Canada through the Department of Innovation, Science and Economic Development Canada and by the Province of Ontario through the Ministry of Colleges and Universities. 

\appendix

\section{Charges of Einstein--Hilbert and metric Chern--Simons theories}

In this appendix we give the Iyer--Wald expressions which we have used in the main text in order to compute the charges \eqref{BS charges} and \eqref{FG charges}. Let us consider the two metric Lagrangians
\be\label{EH and CS Lagrangians}
L=L_\text{EH}=\f{1}{2}\sqrt{-g}\left(R+\f{2}{\ell^2}\right),
\q\q
\check{L}=L_\text{CS}=\f{1}{4}\sqrt{-g}\,\eps^{\mu\nu\rho}\Gamma^\alpha_{\mu\beta}\left(\partial_\nu\Gamma^\beta_{\rho\alpha}+\f{2}{3}\Gamma^\beta_{\nu\gamma}\Gamma^\gamma_{\rho\alpha}\right),
\ee
where $\sqrt{-g}\,\eps^{\mu\nu\rho}=\epsilon^{\mu\nu\rho}$ is the Levi--Civita symbol. Their variations are given by
\be
\delta L=\f{1}{2}\sqrt{-g}\,\delta g^{\mu\nu}\left(G_{\mu\nu}-\f{1}{\ell^2}g_{\mu\nu}\right)+\partial_\mu\Theta^\mu,
\q\q
\delta\check{L}=\f{1}{2}\sqrt{-g}\,\delta g^{\mu\nu}C_{\mu\nu}+\partial_\mu\check{\Theta}^\mu,
\ee
where the Cotton tensor $C_{\mu\nu}$ is defined in terms of the three-dimensional Schouten tensor $S_{\mu\nu}$ as
\be
C_{\mu\nu}={\eps_\mu}^{\rho\sigma}\nabla_\rho S_{\nu\sigma},
\q\q
S_{\mu\nu}=R_{\mu\nu}-\f{1}{4}Rg_{\mu\nu}.
\ee
The symplectic potentials are given by
\be\label{potentials}
\Theta^\mu=\f{1}{2}\sqrt{-g}\,\Big(g^{\alpha\beta}\delta\Gamma^\mu_{\alpha\beta}-g^{\mu\alpha}\delta\Gamma^\beta_{\alpha\beta}\Big),
\q\q
\check{\Theta}^\mu=\f{1}{4}\sqrt{-g}\,\eps^{\mu\nu\rho}\Big(\Gamma^\alpha_{\rho\beta}\delta\Gamma^\beta_{\nu\alpha}-2R^\alpha_\rho\delta g_{\alpha\nu}\Big).
\ee
Under the action of a diffeomorphism we have
\be
\fL_\xi L=\partial_\mu(\xi^\mu L),
\q\q
\fL_\xi\check{L}=\partial_\mu\left(\xi^\mu\check{L}+\f{1}{2}\sqrt{-g}\,\eps^{\mu\nu\rho}\partial_\nu\Gamma^\alpha_{\rho\beta}\partial_\alpha\xi^\beta\right),
\ee
and we can derive the Noether potentials
\be\label{Noether potentials}
K_\xi^{\mu\nu}\coloneqq-\f{1}{2}\sqrt{-g}\,\nabla^{[\mu}\xi^{\nu]},
\q\q
\check{K}_\xi^{\mu\nu}\coloneqq\f{1}{8}\sqrt{-g}\,\eps^{\mu\nu\rho}\Big(\Gamma^\alpha_{\rho\beta}\nabla_\alpha\xi^\beta-4S_{\rho\alpha}\xi^\alpha\Big).
\ee
Note that $\check{L}$ transforms with an anomaly which can be traced back to the transformation behavior $\fL_\xi\Gamma^\lambda_{\mu\nu}=\L_\xi\Gamma^\lambda_{\mu\nu}+\partial_\mu\partial_\nu\xi^\lambda$ of the connection.

Following \cite{Adami:2021sko} and using a convenient normalization factor, we then define the diffeomorphism charges deriving from each Lagrangian as the Iyer--Wald expressions \cite{Iyer:1994ys}
\be\label{charge aspects}
\slashed{\delta}\Q^{\mu\nu}_\xi\coloneqq2\Big(\delta K^{\mu\nu}_\xi-K^{\mu\nu}_{\delta\xi}+\xi^{[\mu}\Theta^{\nu]}\Big),
\q\q
\slashed{\delta}\check{\Q}^{\mu\nu}_\xi\coloneqq2\Big(\delta\check{K}^{\mu\nu}_\xi-\check{K}^{\mu\nu}_{\delta\xi}+\xi^{[\mu}\check{\Theta}^{\nu]}+\Sigma^{\mu\nu}_\xi\Big),
\ee
where the presence of
\be
\Sigma^{\mu\nu}_\xi\coloneqq\f{1}{8}\sqrt{-g}\,\eps^{\mu\nu\rho}\delta\Gamma^\alpha_{\rho\beta}\partial_\alpha\xi^\beta
\ee
can be traced back to the non-covariance of $\check{\Theta}^\mu$. Finally the charge aspects  \eqref{charge aspects} can be written as 
\bsub
\be
\slashed{\delta}\Q^{\mu\nu}_\xi&=\f{1}{2}\sqrt{-g}\left(\xi^\nu\big(\nabla^\mu h-\nabla_\sigma h^{\mu\sigma}\big)+\xi_\sigma\nabla^\nu h^{\mu\sigma}+\f{1}{2}h\nabla^\nu\xi^\mu+h^{\mu\sigma}\nabla_\sigma\xi^\nu\right)-(\mu\leftrightarrow\nu),\\
\slashed{\delta}\check{\Q}^{\mu\nu}_\xi&=\f{1}{2}\sqrt{-g}\,\eps^{\mu\nu\rho}\left(\f{1}{2}\delta\Gamma^\alpha_{\beta\rho}\nabla_\alpha\xi^\beta-\delta S_{\rho\alpha}\xi^\alpha-\f{1}{2}\xi^\beta\big(S^\alpha_\beta h_{\alpha\rho}-S^\alpha_\rho h_{\alpha\beta}\big)\right)-(\mu\leftrightarrow\nu),
\ee
\esub
where the variations are defined as
\be
h_{\mu\nu}\coloneqq\delta g_{\mu\nu},
\q\q
h^{\mu\nu}=g^{\mu\rho}g^{\nu\sigma}h_{\rho\sigma},
\q\q
h\coloneqq g^{\mu\nu}h_{\mu\nu}.
\ee
Note that as an alternative to the Iyer--Wald charges \eqref{charge aspects} we could have defined the charges with the cohomological methods of Barnich and Brandt \cite{Barnich:2001jy}. Both expressions differ by a simple term which can actually be checked to vanish in the present setup. The Barnich--Brandt charges therefore reproduce the results \eqref{BS charges} and \eqref{FG charges} as well.

\bibliography{Biblio.bib}
\bibliographystyle{Biblio}

\end{document}